\documentclass[a4paper,fleqn]{cas-dc}
\usepackage{soul} % 导入 soul 包
\usepackage{color, xcolor} 
\usepackage{hyperref}
\usepackage[numbers]{natbib}
\usepackage[boxed,ruled,linesnumbered]{algorithm2e}

\newtheorem{definition}{Definition}  %gws
\usepackage{subfigure}
\usepackage{tabularx}
\usepackage{soul}
\usepackage{color, xcolor} % 颜色包，color 必须导入，xcolor 建议导入\
\hypersetup{
	colorlinks=blue,  % 链接是否带颜色
	linkcolor=blue,   % 链接颜色
	citecolor=blue,    % 引用颜色
	urlcolor=blue,   % URL链接颜色
}
\sethlcolor{yellow}
\soulregister{\cite}7 % 注册\cite命令
\soulregister{\citep}7 % 注册\citep命令
\soulregister{\citet}7 % 注册\citet命令
\soulregister{\ref}7 % 注册\ref命令
\soulregister{\pageref}7 % 注册\pageref命令
\soulregister{\underline}7 % 注册\underline命令   
\usepackage{float}       % 实现固定位置的[H]选项
\usepackage{threeparttable} % 表格注释关联
\usepackage{array}      
\usepackage{tabularx}
\usepackage[switch]{lineno}   
\usepackage{amsmath}

\begin{document}

\shorttitle{Improve Viral Genomic Feature Representation and Classification} 
\shortauthors{Wenxi Zhu et~al.}	
\title [mode = title]{Mining Negative Sequential Patterns to Improve Viral Genomic Feature Representation and Classification}            
% \tnotetext[1]{This research was partially supported ...}

\author[1]{Wenxi Zhu}
\ead{zhuwenxi317@gmail.com}
\address[1]{College of Information Science and Technology, Jinan University, Guangzhou 510632, China}

\author[2]{Wensheng Gan}
\cortext[cor1]{Corresponding author}
\ead{wsgan001@gmail.com}
\address[2]{College of Cyber Security, Jinan University, Guangzhou 510632, China}
\cormark[1]

\author[3]{Zhenlian Qi}
\ead{qzlhit@gmail.com}
\address[3]{Guangdong Eco-Engineering Polytechnic, Guangzhou 510520, China}

\begin{abstract}
    Viruses represent the most abundant biological entities on Earth and play a pivotal role in microbial ecosystems, yet, as prominent human pathogens, they are closely linked to human morbidity and mortality. Accurate identification of viral sequences from viral genome sequences is therefore essential, but existing genome-based classification models that largely relying on composition- or frequency-based subsequence features often suffer from limited interpretability and reduced accuracy, particularly on complex or imbalanced datasets. To address these limitations, we propose GeneNSPCla (Genomic Negative Sequential Pattern-based Classification), a novel viral classification framework based on Negative Sequential Patterns (NSPs) that extracts discriminative absence-based features from nucleotide sequences of RNA viral genomes. By transforming these NSPs into numerical feature vectors and integrating them into multiple supervised classifiers, GeneNSPCla effectively captures both presence and absence signals in viral sequences. Furthermore, we propose a negative pattern mining algorithm adapted for processing genomic data: GONPM+, which can discover longer and more biologically meaningful negative sequential patterns. The experimental results demonstrate that the average accuracy of GONPM+ in 8 classifiers has improved by 10.03\% compared to the original negative pattern mining algorithm and by 24.75\% compared to the positive pattern mining algorithm. These findings highlight the effectiveness of incorporating absence-based sequential information, providing a new and complementary perspective for viral genome analysis and classification. The source code and datasets are available at https://github.com/zhuwenxi317/GeneNSPCla.
\end{abstract}

\begin{keywords}
    Viral Genomes\\
    Pattern mining\\
    Machine learning classifiers\\ 
    Negative sequential patterns\\
    Classification\\
\end{keywords}

\maketitle

\section{Introduction} \label{sec:introduction}

Viruses, as the most abundant and diverse biological entities on Earth \cite{guerin2018biology}, parasitize various living organisms, exerting profound impacts on ecosystem balance, biological evolution, and human health. On the one hand, some viruses participate in the material cycling of marine plankton and promote horizontal gene transfer in hosts, thereby serving as a crucial driver of ecological evolution \cite{suttle2007marine}. On the other hand, virulent viruses (e.g., SARS-CoV-2, influenza viruses, and Ebola viruses) can trigger zoonotic diseases; they rapidly generate new strains through genomic mutation and recombination, posing a persistent threat to global public health security \cite{tegally2021detection}.  With the rapid advancement of biological sequencing technologies, massive amounts of viral genome data have been continuously emerging, such as GenBank \cite{sayers2019genbank}, and GISAID \cite{burki2023first}, which have significantly promoted the progress of virology research. 

In the field of virology, virus classification \cite{lefkowitz2018virus} is a fundamental task. Accurate virus classification can provide key information for disease diagnosis, epidemic prevention and control, vaccine development, and ecosystem research. Most genomic tools employed to build taxonomies for living organisms rely on alignment-based approaches. Notable examples include BLAST \cite{altschul1990basic} and FASTA \cite{pearson1994using}, as well as their enhanced or expanded variants, which are widely recognized and frequently used as reference standards in genomic sequence analysis \cite{pearson2013blast}. However, alignment-based approaches also have several drawbacks \cite{li2018comparative}. For instance, these methods rely on the sequence collinearity assumption, struggling to handle nonlinear chimeric genomic fragments resulting from high mutation rates, recombination, horizontal gene transfer, and other factors. Second, these models exhibit high computational complexity—algorithms like dynamic programming incur substantial time and memory costs, making them unsuitable with NGS big data \cite{zielezinski2017alignment}. While several alignment-free tools have emerged, such as Kallisto \cite{bray2016near} and Kraken 2 \cite{wood2019improved}, they can handle moderately similar sequences but often falter when applied to highly mutable viruses (e.g., HIV and influenza viruses), as they fail to account for the frequent latent patterns embedded within the sequences \cite{wade2024investigating}.

Against this backdrop, the integration of machine learning (ML) \cite{zhou2021machine} and sequential pattern mining (SPM) techniques \cite{fournier2022pattern,mooney2013sequential} offers a promising approach to overcome traditional bottlenecks. Frequent pattern mining (FPM) \cite{agrawal1993mining} is a type of SPM. FPM can automatically identify high-frequency sequence patterns (such as conserved motifs and functional modules) from large-scale sequence data, while ML algorithms can further transform these patterns into classification features to construct high-accuracy prediction models \cite{remita2017machine}. For example, the GenoAnaCla method significantly improved the precision of viral classification by mining frequent sequence patterns and using machine learning classifiers, demonstrating the effectiveness of pattern mining in sequence analysis \cite{nawaz2024exploiting}. However, current methods based on frequent pattern mining focus on frequent patterns. For example, they predominantly utilize features such as k-mers \cite{liu2024viruspredictor}, frequent sequential patterns of nucleotides/codons/amino acids \cite{nawaz2024exploiting}, and high-frequency CpG islands \cite{gupta2021comparative}. However, this exclusive focus on frequently occurring patterns may overlook another critical dimension of discriminative information: negative patterns. 

In viral genome classification research, negative patterns \cite{wu2004efficient} refer to sequence motifs that are unexpectedly absent or substantially underrepresented within specific viral genomic groups. These absence-based patterns often signal the loss or suppression of critical regulatory or structural elements in certain viral strains, thereby harboring phylogenetic or functional information that is not captured by positive pattern mining alone \cite{georgakopoulos2021absent}. For example, Koulouras et al. \cite{koulouras2021significant} showed that such negative patterns (i.e., minimal absent words) in viral genomes reflect evolutionary negative selection, including the systematic avoidance of host restriction enzyme recognition sites, immune evasion via host-sequence mimicry, and constraints related to genome stability and species-specific regulation. Such patterns offer complementary biological insights into viral genome organization and evolutionary relationships, providing an additional informative dimension for distinguishing viral species and deciphering their adaptive traits. To this end, this paper attempts to present a negative pattern mining-based approach to viral genome classification (GeneNSPCla) from a new perspective. This method aims to provide a complementary dimension of analysis for virus classification by capturing discriminative signals embedded in missing features. The main contributions of this paper are as follows:

\begin{itemize}
    \item The GeneNSPCla method is proposed, which applies negative patterns—defined as patterns that are significantly absent or underrepresented in specific viral taxa—to viral genome classification. This approach offers a way to complement existing methods that primarily rely on frequent patterns.
  
   \item The GONPM+ algorithm is proposed as an extension of ONP-Miner, introducing a series of decay factors to achieve a more flexible and fine-grained dynamic adjustment of the minimum support threshold. This enhancement enables adaptive control of support reduction across different pattern lengths, allowing the algorithm to capture longer negative sequential patterns from RNA viral genomes.
  
    \item Unify the feature formats of positive and negative sequential patterns by standardizing them into a consistent representation to eliminate format differences. Eight machine learning models were used to construct classifiers. Combined with multiple evaluation metrics, a comprehensive analysis of the classification performance of viral data in different formats was performed. In addition, compared with existing mainstream genomic classification methods, the effectiveness and superiority of negative pattern characteristics in virus classification were verified.
\end{itemize}

An initial algorithm was previously introduced in a preliminary version \cite{zhu2025leveraging}. The structure of this paper is organized as follows. Section \ref{sec: relatedwork} provides a summary of related work. Section \ref{sec: Preliminaries} presents the preliminary work of the proposed method, as well as the definitions of relevant terminologies. Section \ref{sec: GeneNSPCla approach} elaborates on the GeneNSPCla approach, encompassing data preprocessing procedures and the specific implementation details of the GONPM+ algorithm. Section \ref{sec: result} presents and discusses the experimental results. Section \ref{sec: conclusion} concludes this study.

\begin{figure*}[ht]
	\centering
	\includegraphics[clip,scale=0.55]{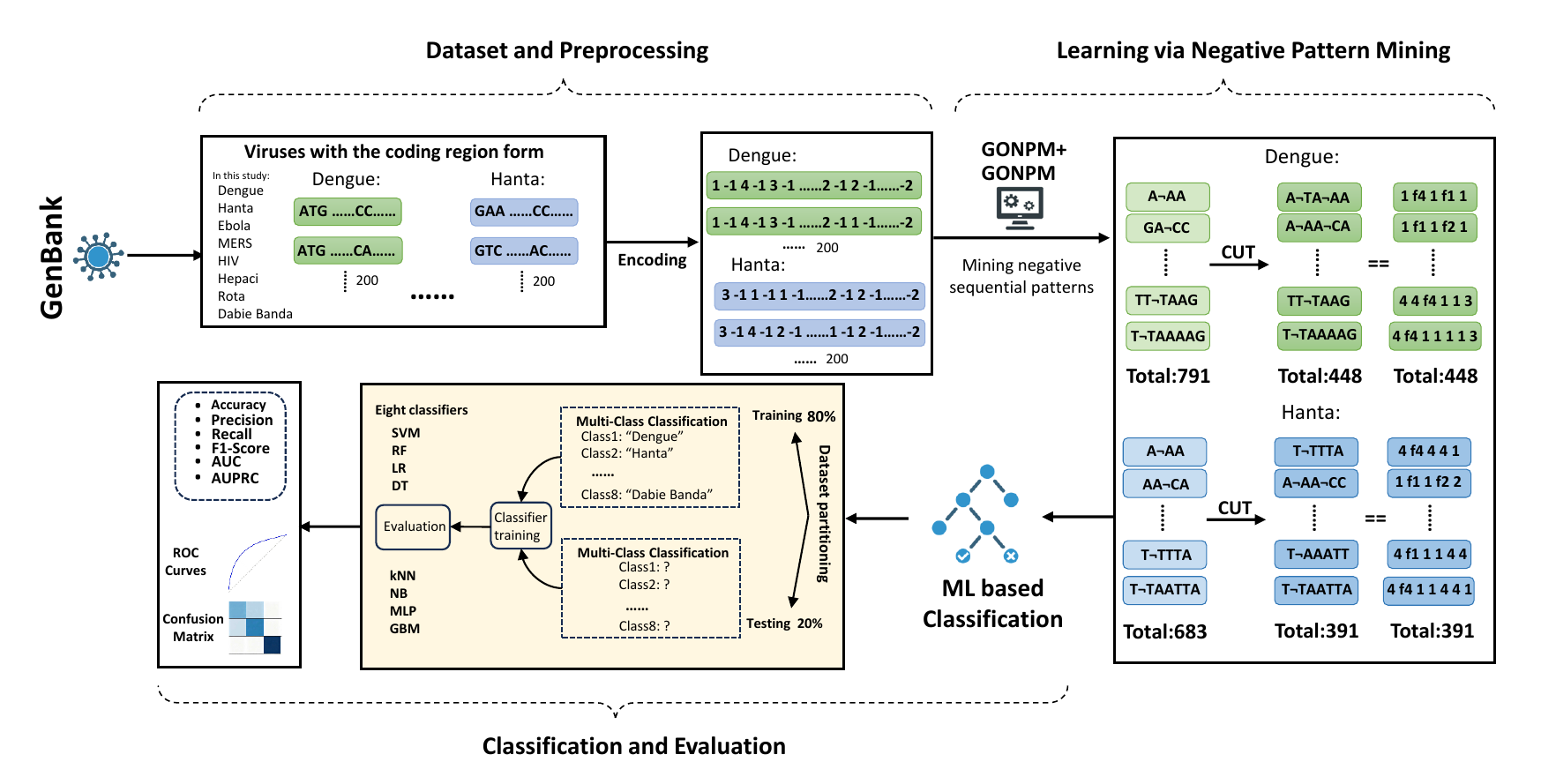}
	\caption{The entire GeneNSPCla framework can be divided into three parts: (1) dataset acquisition and encoding preprocessing; (2) frequent pattern mining via negative pattern algorithms; (3) classification using eight machine learning classifiers with various evaluation metrics obtained. On the right side of the second part is the encoding processing mentioned in (1), with the '-1' symbols between each base and the '-2' symbol at the end omitted.}
	\label{fig:GeneNSPCla}
\end{figure*}

\section{Related Work} \label{sec: relatedwork}

In recent years, computational methods grounded in DL and ML have found widespread application in the analysis, prediction, and classification of genome sequences.

\subsection{ML-based approaches}

Currently, the core logic of ML-based viral classification techniques lies in mining discriminative features from viral genomic data via algorithms and constructing classification models to achieve efficient categorization.  A framework that combines NLP with traditional ML was proposed \cite{alshayeji2023viral}. In this framework, DNA sequences are converted into numerical vectors using various k-mer sizes through the CountVectorizer, with the best classification performance achieved when using 6-mer features. VirusPredictor is an open-source Python software based on XGBoost \cite{chen2016xgboost}, which incorporates the classification of ERVs into viral prediction for the first time. Additionally, this software exhibits a key characteristic: the longer the input sequence, the higher its prediction accuracy \cite{liu2024viruspredictor}. 4CAC is a four-category metagenomic contig classification tool based on ML and assembly graphs. It enables simultaneous identification of viruses, plasmids, prokaryotes, and microeukaryotes, while addressing key limitations of existing classifiers: oversight of class imbalance and low classification accuracy for short contigs \cite{pu20244cac}. The study \cite{ahmed2022enabling} proposed a genomic sequence analysis system for COVID-19 and similar viruses based on the SVM, providing support for viral research and prevention and control at the genomic level. An additional approach utilizes classifiers in ML to classify sublineages of PRRSV. This method addresses the limitations of traditional approaches, specifically the unstable accuracy of RFLP and the high computational cost of phylogenetic analysis \cite{kim2021applications}.

\subsection{DL-based approaches}

In recent years, with the algorithmic innovations and enhanced computing power of DL technology, DL has offered novel perspectives and efficient solutions to the challenge of viral classification, boasting end-to-end features and robust complex data processing capability. Researchers propose a CNN-based viral classification method \cite{gunasekaran2021analysis}. This method directly converts the DNA sequences of six virus types into numerical form via label encoding and K-mer encoding. Besides, three deep learning models, including CNN, CNN-LSTM, and CNN-bidirectional LSTM, were constructed, yielding promising classification performance. The DeepVirusClassifier model is trained based on 1D-CNN, and it combines artificial mutation testing to verify its generalization ability, thus enabling the classification of SARS-CoV-2 and other subtypes of viruses within the Coronaviridae family \cite{azevedo2024deepvirusclassifier}. ViraLM adopts a pretrained genomic foundation model, DNABERT-2, as its backbone. By encoding input sequences after splitting them into 2 kb segments, it enables novel virus identification in metagenomes using average prediction scores \cite{peng2024viralm}. DNASimCLR processes DNA sequences using One-Hot encoding, utilizes the SimCLR framework and ResNet-50 encoder for training on unlabeled data, and achieves better classification performance than existing methods when tested on three benchmark datasets, including the virus-host dataset \cite{yang2024dnasimclr}. DeepMicroClass employs a dual-path CNN architecture, which comprises a base path (via One-Hot encoding) and a codon path (via three reading frame conversion). When tested on 20 synthetic benchmark datasets, this model achieved promising performance in classifying eukaryotic, plasmid, and viral sequences \cite{hou2024deepmicroclass}.

\subsection{Feature mining approaches}

In viral classification, feature mining serves as the core link connecting raw viral data and classification models.  Its objective is to extract discriminative key information from genomic sequences.  Classic feature extraction methods involve extracting k-mer frequency features.  For example, VirFinder, the first ML tool based on k-mer frequencies, uses k-mer signals in full sequence to identify viral prokaryotic sequences from assembled metagenomic data \cite{ren2017virfinder}.  Previous research uses three feature encoding algorithms—amino acid composition (AAC), parallel correlation pseudo-amino acid composition (PC-PseAAC), and G-gap dipeptide composition (GGAP)—to process the spike protein sequences of coronaviruses.  The processed sequences generated feature values, which were used for training the random forest (RF) model \cite{qiang2020using}.  Association rule mining (ARM) has been applied to extract symptom association rules, helping to identify symptom patterns in COVID-19 patients \cite{tandan2021discovering}.  GenoAnaCla processes 15 types of viral RNA sequences, employing the CM-SPAM algorithm \cite{fournier2014fast} to extract frequent sequence patterns. These outputs are ultimately fed into ML classifiers, achieving promising performance \cite{nawaz2024exploiting}.  SPM4GAC employs two SPM algorithms, namely CM-SPAM and TKS, to extract frequent nucleotide sequence patterns composed of the four canonical nucleotides (A, C, G, T) as features \cite{nawaz2024spm4gac}.  FSP4HSP \cite{nawaz2024fsp4hsp} employs frequent pattern mining algorithms to extract Frequent Sequential Patterns of Amino Acids (FSPAAs) and mines the sequential rules between amino acids via the ERMiner algorithm.  In addition, there exist related methods that leverage the DALI tool to identify similar protein structures in the Protein Data Bank (PDB) and employ these structural elements as classification features \cite{nawaz2022s,nawaz2023psac}.  For extracting biomarkers based on Digital Signal Processing (DSP) technology, the method is grounded in the tri-nucleotide periodicity of DNA and employs DSP technology to process viral sequences before inputting the processed data into ML classifiers for classification \cite{singh2021classification}. Furthermore, there is a nonlinear feature extraction method based on chaos game representation (CGR) and recurrence quantification analysis (RQA) \cite{olyaee2020rcovid19}.

\subsection{Challenges in existing approaches}

Despite the progress achieved by ML- and DL-based approaches in viral classification, several limitations remain. For ML-based methods, performance largely depends on the quality of manually engineered features such as k-mer counts. Although larger k-mer sizes generally lead to higher classification accuracy \cite{randhawa2020machine}, this improvement comes at the cost of exponentially increasing memory and computational requirements. Specifically, the memory complexity of traditional k-mer methods is $(4^k)$, making it challenging for standard computing devices to handle large k values \cite{shi2020general,moeckel2024survey}. Moreover, such methods exhibit limited discriminative power for closely related viral strains with high genetic similarity and may suffer from reduced classification accuracy when highly similar genomes are excluded \cite{akbari2023new}.

DL-based methods, although capable of automatic feature extraction, face challenges including high computational cost, the need for large labeled datasets, and limited interpretability of the learned representations. For instance, traditional DL models exhibit significantly pronounced "black-box" characteristics and poor interpretability. They thus rely on complex and time-consuming auxiliary methods, such as DeepLIFT attribution combined with TF-MoDISco clustering, filter visualization, and filter invalidation—to deduce biological significance \cite{novakovsky2023explainn,przymus2025deep}. Additionally, there are significant issues at the data level: on the one hand, data scarcity and class imbalance lead to problems such as model bias toward the majority class and poor generalization ability \cite{salmi2024handling}; on the other hand, data annotation requires completion by professional medical personnel and is subject to regulatory restrictions on medical data sharing across different countries, resulting in issues of difficult verification of annotation accuracy and low annotation efficiency \cite{islam2024challenges}. As for feature mining approaches, conventional techniques such as frequent k-mer \cite{ren2017virfinder} extraction, association rule mining \cite{tandan2021discovering}, or chaos game representation \cite{almeida2001analysis} often focus only on positive or frequent patterns, thereby overlooking potentially informative absence signals and rare but discriminative motifs. In fact, studies have shown that relying solely on positive sequence patterns can lead to biased or incomplete similarity and classification results. Incorporating negative sequence patterns, which capture the absence of specific subsequences, provides complementary information that enhances discrimination power \cite{li2025mining}.

To address these limitations, our work explores a complementary perspective by leveraging NSPs to characterize absence-based information in viral genomes. Rather than replacing conventional positive pattern mining, the proposed approach extends positive pattern representations by incorporating negative sequential patterns, which retain information derived from frequent patterns while additionally encoding the absence or underrepresentation of specific elements within viral taxa. This joint representation enriches the feature space used for classification and provides complementary information beyond presence-based patterns alone. Compared with those resource-intensive algorithms that rely on large language model training \cite{wu2023multimodal,zheng2025large}, this data-mining–based framework, combined with conventional machine learning classifiers, offers a lightweight alternative under the evaluated experimental settings while maintaining competitive classification performance.

\section{Preliminaries} \label{sec: Preliminaries}

Sequential pattern mining \cite{fournier2017survey}, a pivotal approach in knowledge discovery, aims to identify meaningful subsequences that are referred to as patterns within individual sequences or sequence databases (SDBs). Over the years, a wide range of SPM algorithms have been proposed, such as SPM with gap constraints \cite{wu2021hanp,wu2021top} and utility-driven SPM \cite{gan2021survey,hossain2021hsnp}. However, most of these methods primarily focus on mining events that have occurred, referred to as positive SPM \cite{fournier2014fast,mordvanyuk2022vepreco}. In contrast, negative SPM emphasizes events that were expected but did not occur. Such negative patterns can be particularly informative, with applications in diverse fields including behavior analysis, medical services, financial risk management, and fraud detection. For instance, SN-RNSP \cite{sun2024sn} has been proposed to extract frequent negative patterns from transaction sequences, while algorithms such as e-RNSP \cite{dong2018rnsp}, NegPSpan \cite{guyet2020negpspan}  have also been developed to mine frequent negative patterns across different domains. In the present study, we focus primarily on negative SPM algorithms suitable for processing large-scale biological datasets. To provide a formal explanation of our algorithm, we first introduce the fundamental concepts commonly used in negative SPM.

\begin{definition}[subsequences \cite{agrawal1995mining}]
    \rm Let \textbf{S} = $\langle$$s_1$, $s_2$, $\dots$, $s_m$$\rangle$ be a sequence. A sequence $\mathbf{S_i}$ = $\langle$$s_{i_1}$, $s_{i_2}$, $\dots$, $s_{i_n}$$\rangle$ with $1 \leq$ $i_1 < i_2 < \cdots < i_n$ $\leq m$ and $n \leq m$ is defined as a subsequence of $\textbf{S}$. When $n$ = $m$ and $i_j$ = $j$ for $j$ = $1$, $\ldots$, $m$, $\mathbf{S_i}$ is identical to $\textbf{S}$. We use the notation $\mathbf{S_i} \sqsubseteq \textbf{S}$, where the symbol $\sqsubseteq$ represents the subsequence relationship. 
\end{definition}

For example, consider the sequence $\langle \{\text{GTTCAACTG}\} \rangle$. Since the elements of $\langle \{\text{TCA}\} \rangle$ appear in the same order within $\langle \{\text{GTTCAACTG}\} \rangle$, $\langle \{\text{TCA}\} \rangle$ is a subsequence of $\langle \{\text{GTTCAACTG}\} \rangle$, which can be denoted as $\langle \{\text{TCA}\} \rangle \sqsubseteq \langle \{\text{GTTCAACTG}\} \rangle$.

\begin{definition}[gap constant \cite{pei2007constraint}]
    \label{def:Gap Constant}
   \rm  A sequence \textbf{S} = $\langle$$s_1$, $s_2$, $\dots$, $s_i$, $\dots$ $s_m$$\rangle$ over an alphabet $\Sigma$ has length $m$, where $s_i \in \Sigma$ for $1 \leq i \leq m$. A pattern \textbf{p} with gap constraints is defined as $\textbf{p}$ = $p_1[M,N]p_2\ldots [M,N]p_j\dots[M,N]p_n$, where $n$ is the length of the pattern, $0 \leq M \leq N$, and $p_j \in \Sigma$. The gap constants $M$ and $N$ specify the minimum and maximum number of wildcard characters allowed between consecutive pattern elements $p_{j-1}$ and $p_j$.
\end{definition}

For example, in an \textit{RF} sequence $\textbf{S}$ = $\langle \{\text{ATTACG}\} \rangle$, the character set $\Sigma$ = $\{\text{A}, \text{G}, \text{C}, \text{T}\}$ and the length $m$ = $6$. Suppose that we have a pattern with gap constraints \textbf{p} = $\text{A}[1,3]\text{C}$, which indicates that between the 'A' and 'C' in the pattern, there can be at least 1 and at most 3 wildcards. Sequences like $\text{AC}$, $\text{ATC}$, $\text{ATTC}$ could potentially match this pattern, depending on how the wildcards are interpreted in the search context.

\begin{definition}[One-off condition \cite{wu2023onp}]
    \rm Suppose $\mathbf{S_1}$ = $\langle$$s_1$, $s_2$, $\ldots$, $s_k$, $\ldots$, $s_m$ $\rangle$ and $\mathbf{S_1'}$ = $\langle$$s'_1$, $s'_2$, $\ldots$, $s'_j$, $\ldots$, $s'_m$$\rangle$ are two occurrences in sequence $\mathbf{S}$. If $s_k$ = $s'_j$ ($1 \leq k \leq m$, $1 \leq j \leq m$), $\mathbf{S_1}$ and $\mathbf{S_1'}$ violate the one-off condition. Otherwise, the two occurrences are said to satisfy the one-off condition.
\end{definition}

Suppose that we have a sequence $\mathbf{S}$ = $\langle\{\text{AACACCTC}\}\rangle$ and a pattern $\mathbf{p}$ = $\text{A}[0,1]\text{C}[0,1]\text{C}$. According to the gap constraint $[0,1]$, all occurrences of $\mathbf{p}$ in $\mathbf{S}$ are $\langle1$,$3$,$5\rangle$, $\langle2$,$3$,$5\rangle$, $\langle4$,$5$,$6\rangle$ and $\langle4$,$6$,$8\rangle$. The occurrences $\langle1$,$3$,$5\rangle$ and $\langle4$,$5$,$6\rangle$ do not satisfy the one-off condition since they share position $5$, whereas $\langle1$,$3$,$5\rangle$ and $\langle4$,$6$,$8\rangle$ do satisfy it.

\begin{definition}[negative pattern \cite{wu2023onp}]
    \rm Give a character $ e $ (where $ e \in \Sigma $ or $ e $ is the null character), the corresponding negative character is denoted as $ \neg e $, and this symbol represents the absence of the character $ e $. A negative pattern that incorporates gap constraints can be formulated as:
\[
\mathbf{p} = p_1[M,N]\neg e_1 p_2 \cdots [M,N]\neg e_{j-1} p_j \cdots [M,N]\neg e_{m-1} p_m \quad \](1) If $e_{j-1}$ $\in$ $\Sigma$, then $p_{j-1}[M,N]$ $\neg$ $e_{j-1} p_j$ is a negative subsequence, which means that $e_{j-1}$ does not exist between $p_{j-1}$ and $p_j$. (2) If $e_{j-1}$ is $\text{null}$, then $p_{j-1}[M,N] p_j$ is a classical example in \hyperref[def:Gap Constant]{Definition 2}.
\end{definition}

For example, consider a pattern \textbf{p} = $p_1[0,1]p_2[0,2]$$\neg$$e_1$$p_3$ = $\text{C}[0,1]\text{A}[0,2]\neg \text{GT}$. Here, the negative character $\neg \text{G}$ in the subpattern $\text{A}[0,2]\neg \text{GT}$ indicates that between "A" and "T", there can be 0, 1, or 2 characters of any type except "G"—in other words, "G" must not appear in the interval between "A" and "T".

\begin{definition}[minimum support \cite{wu2023onp}]
    \rm The support of a pattern \textbf{p} within a sequence \textbf{S} is defined to be the count of its occurrences subject to the one-off constraint, represented as \textit{sup}(\textbf{p}, \textbf{S}). For a sequence dataset  \textit{SDB}, the minimum support of a pattern \textbf{p} is \textit{sup}(\textbf{p}, \textit{SDB}), calculated as: \textit{sup}(\textbf{p}, \textit{SDB}) = $\sum_{k=1}^{N}$ \textit{sup}(\textit{p}, $s_k$) ,where \textit{N} is the number of sequences in \textit{SDB}. If the support of pattern \textbf{p} in \textit{SDB} is greater than or equal to a predefined threshold \textit{minsup}, that is, \textit{sup}(\textbf{p}, \textit{SDB}) $\geq$ \textit{minsup}, then \textbf{p} is referred to as a frequent sequence.
\end{definition}

Note that the concept of support calculation here differs from that in traditional SPM algorithms. In conventional SPM algorithms, if a pattern $\textbf{p}$ appears in a sequence \textbf{S}, its support is counted as $1$ regardless of the number of occurrences. In contrast, in the GONPM+ algorithm, the support is determined by the total number of times \textbf{p} appears across all sequences. For example, given the sequence \textbf{S} = $\langle$$\{\text{AACACACCTC}\}$$\rangle$ and the pattern \textbf{p} = $\text{A}$$[0,1]$$\text{C}$$[0,1]$$\neg$\text{GC}, $\textbf{p}$ can match both the prefix $\text{'AACAC'}$ and the suffix $\text{'ACCTC'}$ of $\textbf{S}$. Therefore, the support of \textbf{p} in \textbf{S} is $2$.  Based on the concept of subsequence, the support measure is defined as follows. \textit{GSC} refers to the genome sequence corpus and $S_x$ is a subsequence of a certain sequence $S$. The support of the subsequence $S_x$, denoted as \textit{sup}($S_x$), is the total occurrence frequency of the subsequence $S_x$ in \textit{GSC}. \textit{Count}($S_x$, $S$) denotes the number of occurrences of $S_x$ in $S$. Formally, it is defined as:
\begin{equation}
    \textit{sup}(S_x) = \sum_{S \in GSC} \textit{count}(S_x, S).
\end{equation}

\section{Proposed GeneNSPCla approach}  \label{sec: GeneNSPCla approach}

In this section, we introduce the GeneNSPCla method. The GeneNSPCla method is divided into three main steps: 1) Database and processing; 2) Obtaining negative sequential patterns of data through the GONPM+ algorithm; and 3) Training classifiers using these patterns and performing performance evaluation. The GeneNSPCla method can serve as a general approach, allowing the configuration of different negative SPM algorithms or distinct classifiers. 

\subsection{Database and preprocessing}
\label{def:Database and processing}

\newcolumntype{L}{>{\raggedright\arraybackslash}X} % 左对齐的自适应列
\newcolumntype{R}{>{\raggedleft\arraybackslash}X}  % 右对齐的自适应列

\begin{table}[ht]
\centering
\footnotesize
\begin{threeparttable}
\caption{Viral genome sequences were retrieved from the NCBI GenBank database in the CRF, and additional statistics on their lengths were conducted.}
\label{tab:Table1}
\begin{tabular}{p{1cm} c c c c} % 列格式：左对齐+左对齐+右对齐×4
\hline
Virus       & RNA type          & Samples in CRF & minlen & maxlen \\ \hline
Dengue      & (+)ssRNA          & 200           & 8136   & 10179  \\
Dabie       & ($-$)ssRNA        & 200           & 882  & 6255     \\
Hanta       & ($-$)ssRNA        & 200           & 324  & 6477   \\

Ebola       & ($-$)ssRNA        & 200            & 12486  & 16956  \\
MERS        & (+)ssRNA          & 200            & 8568  & 44097   \\
HIV         & ($-$RT)ssRNA      & 200           & 1794   & 11304    \\
Hepaci      & (+)ssRNA          & 200           & 5967   & 10293    \\

Rota        & dsRNA             & 200           & 453  & 3507   \\
 \hline
Average       &                 & 200           & 4826   & 13633    \\ \hline
\end{tabular}
\parbox{\linewidth}{
    (+)ssRNA: Positive-sense single-stranded RNA, ($-$)ssRNA: Negative-sense single-stranded RNA, dsRNA: Double-stranded RNA. ($-$RT)ssRNA: Negative-sense reverse transcription single-stranded RNA, CRF: Coding Region Form.}
\end{threeparttable}
\end{table}

We selected eight RNA viruses from distinct taxonomic groups that exhibit low pairwise similarity and considerable variation in sequence length, with sequencing data retrieved from GenBank \cite{sayers2019genbank}. For each virus, we focused on the \textit{coding region form (CRF)}, which represents continuous coding regions composed of codons, where each codon consists of three nucleotides. These CRF sequences were downloaded in FASTA format \cite{pearson1994using}. Notably, the GeneNSPCla framework is representation-agnostic and can be extended beyond CRF to non-coding regions or protein-level amino acid sequences in our future studies. Due to the high computational complexity of extracting negative sequential patterns, the dataset used for GONPM training was constructed by randomly selecting 200 genomic sequences from each of the eight RNA virus species. This sampling strategy ensures that each virus type contributes an equal number of sequences, thus mitigating potential data imbalance between categories and maintaining diversity within each class. However, the current version of GONPM+ is not yet optimized for large-scale datasets, as mining negative patterns over the entire genomic collection would result in substantial memory consumption and computational overhead. We expect that training on larger datasets would further enhance the robustness and generalizability of the discovered negative patterns, providing a more comprehensive representation of viral genomic diversity once computational efficiency is improved. Future work will therefore focus on training the model with larger datasets and evaluating its performance on closely related viral strains as well as real-world metagenomic mixtures. The detailed composition of the dataset is summarized in Table \ref{tab:Table1}.

Each viral genome sequence consists of four canonical nucleotides: \textit{A} (adenine), \textit{C} (cytosine), \textit{G} (guanine), and \textit{T} (thymine). In some sequences, additional symbols appear to represent ambiguous positions or mixtures of these bases; such cases are referred to as \textit{redundant nucleotides (RN)}. Samples containing RN are thus defined as redundant samples, and in our dataset, the proportion of redundant sequences was approximately 6.47\%. To handle these, each distinct RN was mapped to a unique positive integer. In this encoding scheme, a special code $-1$ serves as a delimiter between codons, and $-2$ is appended at the end of each sequence to denote termination. This conversion enables direct processing by conventional SPM algorithms, thereby ensuring compatibility and allowing seamless comparison with GONPM. Table~\ref{table:genome_datasets} illustrates CRF sequences in both raw and encoded formats, with examples covering positive and negative SPM. 

Following this encoding scheme, for each virus class, sequences were encoded as illustrated in Table \ref{table:genome_datasets}, which served as input to the positive frequent pattern mining algorithm CM-SPAM. In contrast, the original genomic sequences were directly used as input for the negative pattern mining algorithm GONPM+ without applying additional sequence filtering or length normalization, since GONPM+ operates on complete sequence information. The discovered negative sequential patterns were subsequently encoded using the same scheme as positive patterns, thereby unifying both representations into a consistent feature format for downstream classification.

\begin{table*}[ht]
    \caption{(CRF)-formatted genome sequences, including five examples. The original sequences are presented in Figure (a); Figure (b) presents the encoded sequences; Figure (c) displays a part of the negative sequential patterns corresponding to each example; and Figure (d) shows the encoding of the negative sequential patterns in Figure (c).}
    \label{table:genome_datasets}
    \centering
    \small
\begin{tabular}{|c|c|}
    \hline
    \multicolumn{2}{|c|}{\textbf{(a) Positive sequences in \textit{CRF}}} \\
    \hline
    \textbf{ID} & \textbf{Sequence} \\
    \hline\hline
    1 & $\langle \{ \text{AGC} \} \rangle$ \\
    \hline
    2 & $\langle \{ \text{T[0,2]ACG} \} \rangle$ \\
    \hline
    3 & $\langle \{ \text{TA[0,2]GC[0,2]TA} \} \rangle$ \\
    \hline
    4 & $\langle \{ \text{T[0,2]ATGC[0,2]AT} \} \rangle$ \\
    \hline
    5 & $\langle \{ \text{GC[0,2]G[0,2]TA[0,2]CG} \} \rangle$ \\
    \hline
\end{tabular}
\quad
\begin{tabular}{|c|c|}
    \hline
    \multicolumn{2}{|c|}{\textbf{(b) The transformed sequences in \textit{CRF}}} \\
    \hline
    \textbf{ID} & \textbf{Sequence} \\
    \hline\hline

    \hline
    1 & $\mathbf{1}$ -- $1$ $\mathbf{3}$ -- $1$ $\mathbf{2}$ -- $1$ -- $2$ \\
    \hline
    2 & $\mathbf{4}$ -- $1$ $\mathbf{1}$ -- $1$ $\mathbf{2}$ -- $1$ $\mathbf{3}$ -- $1$ -- $2$ \\
    \hline
    3 & $\mathbf{4}$ -- $1$ $\mathbf{1}$ -- $1$ $\mathbf{3}$ -- $1$ $\mathbf{2}$ -- $1$ $\mathbf{4}$ -- $1$ $\mathbf{1}$ -- $1$ -- $2$ \\
    \hline
    4 & $\mathbf{4}$ -- $1$ $\mathbf{1}$ -- $1$ $\mathbf{4}$ -- $1$ $\mathbf{3}$ -- $1$ $\mathbf{2}$ -- $1$ $\mathbf{1}$ -- $1$ $\mathbf{4}$ -- $1$ -- $2$ \\
    \hline
    5 & $\mathbf{3}$ -- $1$ $\mathbf{2}$ -- $1$ $\mathbf{3}$ -- $1$ $\mathbf{4}$ -- $1$ $\mathbf{1}$ -- $1$ $\mathbf{2}$ -- $1$ $\mathbf{3}$ -- $1$ -- $2$ \\
    \hline
\end{tabular}

\vspace{1em}
    
\begin{tabular}{|c|c|}
    \hline
    \multicolumn{2}{|c|}{\textbf{(c) Negative sequences in \textit{CRF}}} \\
    \hline
    \textbf{ID} & \textbf{Sequence} \\
    \hline\hline
    1 & $\langle \{ AT[0,2]\neg G \} \rangle$ \\
    \hline
    2 & $\langle \{ A[0,2]TG[0,2]\neg C \} \rangle$ \\
    \hline
    3 & $\langle \{ GCTA[0,2]\neg T[0,2]\neg G \} \rangle$ \\
    \hline
    4 & $\langle \{ T[0,2]\neg G[0,2]AC[0,2]\neg A \} \rangle$ \\
    \hline
    5 & $\langle \{ A[0,2]\neg CT[0,2]TG[0,2]\neg T \} \rangle$ \\
    \hline
\end{tabular}
\quad
\begin{tabular}{|c|c|}
    \hline
    \multicolumn{2}{|c|}{\textbf{(d) The transformed sequences in \textit{CRF}}} \\
    \hline
    \textbf{ID} & \textbf{Sequence} \\
    \hline\hline
    1 & $\mathbf{1}$ -- $1$ $\mathbf{4}$ -- $1$ $\mathbf{f3}$ -- $1$ -- $2$ \\
    \hline
    2 & $\mathbf{1}$ -- $1$ $\mathbf{4}$ -- $1$ $\mathbf{3}$ -- $1$ $\mathbf{f2}$ -- $1$ -- $2$ \\
    \hline
    3 & $\mathbf{3}$ -- $1$ $\mathbf{2}$ -- $1$ $\mathbf{4}$ -- $1$ $\mathbf{1}$ -- $1$ $\mathbf{f4}$ -- $1$ $\mathbf{f3}$ -- $1$ -- $2$ \\
    \hline
    4 & $\mathbf{4}$ -- $1$ $\mathbf{f3}$ -- $1$ $\mathbf{1}$ -- $1$ $\mathbf{2}$ -- $1$ $\mathbf{f1}$ -- $1$ -- $2$ \\
    \hline
    5 & $\mathbf{1}$ -- $1$ $\mathbf{f2}$ -- $1$ $\mathbf{4}$ -- $1$ $\mathbf{4}$ -- $1$ $\mathbf{3}$ -- $1$ $\mathbf{f4}$ -- $1$ -- $2$ \\
    \hline
\end{tabular}
\end{table*}

\subsection{Using negative SPM for learning}

NSP mining has attracted attention in bioinformatics because it can capture informative absence signals that are often ignored by conventional frequent-pattern mining. The current research on negative SPM mainly focuses on multiple supports \cite{xu2017msnsp}, progressive pattern mining \cite{huang2020mining}, and so on. However, these algorithms exhibit two key limitations: overlooking pattern repeat counting (i.e., only determining whether a pattern is present, without counting its multiple occurrences within a sequence) and a lack of gap constraints. The frequency of a pattern within a single sequence is of crucial importance. For instance, the occurrence frequency of short fragments such as dinucleotides in a single genomic sequence can be applied to species classification and evolutionary research \cite{kariin1995dinucleotide}. Furthermore, due to a lack of gap constraints, these traditional methods incorporate irrelevant regions into patterns during the mining process, resulting in generated patterns containing non-functional fragments, which are essentially meaningless patterns \cite{liao2013efficient}.

\subsubsection{ONP-Miner algorithm}

By contrast, ONP-Miner \cite{wu2023onp} introduces the concept of one-off negative sequential patterns with embedding constraints, thereby reducing redundancy and improving efficiency. Its pruning strategies make it better suited for large sequential datasets than earlier NSP algorithms. Our proposed algorithm adopts ONP-Miner as the foundation and proposes the GONPM+. Our improved algorithm not only preserves the computational efficiency of ONP-Miner but also adapts its definitions of patterns and constraints to the properties of nucleotide and coding-region sequences. This enables the discovery of negative patterns that correspond more closely to biologically meaningful motifs, while maintaining scalability for long viral genomes. The pseudocode of the ONP-Miner (Algorithm \ref{alg: ONP-Miner}) is outlined as follows:

\begin{algorithm}[ht]
    \small
    \caption{ONP-Miner}
    \label{alg: ONP-Miner}
    \LinesNumbered
    \KwIn{A sequence database \textit{SDB}, a minimum support threshold \textit{minsup}, the gap constraint \textit{gap}.} 
    \KwOut{ONPs stored in $F$.}
    
    initialize traverse \textit{SDB}, get $\Sigma$, and store frequent positive patterns with length one in $F_1$; \textit{len} $\gets$ 2;
    
    initialize $F_2$ $\gets$ \textbf{FindONP2} \textbf{(\textit{SDB}, \textit{minsup}, \textit{gap}, $F_1$, $\Sigma$)};    
    
    \While{\rm $F_{len}$$ \neq$ \text{null}}{
         \textit{cand} $\gets$ \textbf{PatternJoin($F_{len}$, $len$)}; \tcp{\upshape\rmfamily Generate all candidate patterns with length} \textbf{\textit{len} + 1;}
                
        \For{\rm each positive candidate pattern \textbf{p} in \textit{cand} }{
         \textit{sup}(\textbf{p}, \textit{SDB}) $\gets$ \textbf{MatchDB(\textit{SDB}, \textbf{p})} ;
        
            \If{\rm  \textit{sup}(\textbf{p}, \textit{SDB}) $\geq$ \textit{minsup} }{
                $ F_{len + 1}$$ \gets$ $F_{len + 1}$$ \cup$  \textbf{p};
                
                Prune \textbf{p};
            }
            
            \Else{
                prune \textbf{p} and its corresponding negative sequence patterns;
            }
        }
         $F_{len + 1}$$ \gets$ $F_{len + 1}$$ \cup$ \textbf{FindFrequent(\textit{SDB}, \textit{minsup}, \textit{cand})};
        
        \textit{len} $\gets$ \textit{len} + 1;
    }
    \Return $F$
\end{algorithm}

\begin{description}
    \item[Step 1:] Traverse the sequence database to identify the frequent patterns of length-1 and store them in $F_1$.
    
    \item[Step 2:] The FindONP2 algorithm is used to generate positive candidate patterns of length 2. It then prunes the infrequent candidate patterns. Finally, it generates negative candidate patterns based on the frequent patterns and calculates the support of each pattern to screen out frequent patterns of length 2. In the FindONP2 algorithm, the FindFrequent algorithm is invoked. The primary function of this algorithm is to filter the set of candidate input patterns and retain frequent patterns with support $\geq$ \textit{minsup}.
    
    \item[Step 3:] The PatternJoin algorithm generates all candidate patterns of length \textit{len} + $1$. Based on the frequent pattern set of length $l$, this algorithm extracts the suffix (by removing the first element) and prefix (by removing the last element) for each pattern in the set. If the suffix of one pattern matches the prefix of another pattern exactly, a candidate pattern of length $l$ + $1$ is generated through "pattern join ($ p \oplus q $)".
    
    \item[Step 4:] The MatchDB algorithm calculates the support of each positive candidate pattern. Centered on depth-first search combined with backtracking, this algorithm relies on the DFS algorithm. It locates unused characters in the sequence that match the first element of the pattern as roots, and through the DFS algorithm, identifies complete occurrences of the pattern that satisfy the one-off, gap, and negative character constraints. After marking the used characters, it backtracks to find new occurrences and finally counts the support of the pattern. This algorithm processes the candidate patterns: if the support of a candidate pattern is no less than the preset threshold, the algorithm stores the pattern in $F_{len + 1}$; otherwise, it prunes the corresponding negative candidate patterns of the candidate pattern.
    
    \item[Step 5:] The MatchDB algorithm computes the support of each negative candidate pattern. If the support of the negative candidate pattern meets or exceeds the predefined threshold, the algorithm stores this pattern in $F_{len+1}$. At this point, $F_{len+1}$ contains both positive and negative patterns with support greater than the threshold, and then \textit{len} is updated to \textit{len} + 1.
    
    \item[Step 6:] Repeat Steps 3 through 5 until the collection of frequent patterns becomes empty.
\end{description}

\subsubsection{GONPM+ algorithm}

To extract discriminative features from RNA viral genomic sequences, we employ a negative sequential pattern mining approach that captures informative absence signals specific to RNA viruses. Building upon the ONP-Miner framework, we previously proposed the GONPM algorithm, which dynamically adjusts the minimum support threshold to enhance the extraction of biologically meaningful negative sequential patterns. This method demonstrated superior performance in identifying informative negative features and effectively improved viral classification accuracy.

In this extended version, we further enhance GONPM by introducing a decay-based adaptive support mechanism, referred to as \textbf{GONPM+}. Unlike the original GONPM, which only adjusts the minimum support once after a specific iteration, GONPM+ introduces a set of decay factors \{$f_2$, $f_3$, $\dots$, $f_n\}$ that progressively reduce the support threshold across multiple levels of candidate generation. This gradual adjustment allows the algorithm to relax the mining constraints in a layer-wise manner, thereby facilitating the exploration of longer and less frequent negative patterns while avoiding excessive candidate expansion. As a result, the mining process becomes more flexible and fine-grained, enabling a more detailed characterization of variations among RNA viral genomes. Consequently, we have made the following improvements:

\begin{itemize}
    \item \textbf{Optimization for adaptability to genomic data}: The storage format of patterns in \(F_{len + 1}\) is optimized and encoded to improve compatibility with genomic data processing (encoded in the format mentioned in Section \ref{def:Database and processing}). Additionally, the algorithm's output results are enhanced to facilitate the direct extraction of processed negative pattern sequences.

    \item \textbf{Progressive decay-based adjustment of the minimum support threshold}: Building upon the adaptive threshold mechanism introduced in GONPM, the improved algorithm GONPM+ incorporates a multi-level decay strategy to further optimize pattern extraction. Instead of applying a fixed reduction after a specific iteration, GONPM+ introduces a series of decay factors $\{f_2, f_3, \dots, f_n\}$ to progressively decrease the \textit{minsup} value across successive invocations of the PatternJoin algorithm. This progressive adjustment enables a smoother and more adaptive relaxation of the support threshold as the pattern length increases, facilitating the discovery of longer and rarer negative patterns while maintaining computational stability and efficiency.
\end{itemize}

The GONPM+ algorithm is shown in Algorithm \ref{alg:GONPM+}. The specific steps of this algorithm are as follows:
\begin{description}
    \item[Step 1:] Traverse the sequence database to identify all frequent length-1 patterns and store them in $F_1$. For the RNA viral genomes analyzed in this study, $F_1$ includes the four canonical nucleotides (\textit{A}, \textit{C}, \textit{G}, and \textit{T}) along with a small number of low-frequency redundant nucleotides. Subsequently, the algorithm invokes the FindONP2 algorithm to generate positive candidate patterns of length 2.
    
    \item[Step 2:] The PatternJoin algorithm is invoked to generate all candidate patterns of length $\textit{len}+1$. For layer-wise control, GONPM+ applies decay factors $(f_2, f_3, \dots, f_n)$ to the original minimum support \textit{minsup}. Specifically, the patterns stored in $F_2$ use a threshold of $\textit{minsup}\times f_2$, the patterns in $F_3$ use $\textit{minsup}\times f_3$, etc., up to $F_n$, which uses $\textit{minsup}\times f_n$. For any subsequent layers beyond $n$, the threshold remains at $\textit{minsup}\times f_n$. 
    
    \item[Step 3:] The MatchDB algorithm is used to calculate the support of each positive candidate pattern. If the support of a candidate pattern meets or exceeds the preset threshold, the pattern is stored in $F_{len + 1}$. Otherwise, the corresponding negative candidate patterns associated with it are pruned in this step of the process.
    
    \item[Step 4:] The MatchDB algorithm is then used to compute the support of each negative candidate pattern. Negative patterns whose support values meet or exceed the predefined threshold are retained and stored in $F_{\text{len} + 1}$ for subsequent iterations. Specifically, $F_{len + 1}$ stores both frequent positive and negative patterns, and \textit{len} is updated to \textit{len} + 1.
    
    \item[Step 5:] Repeat Steps 2 through 4 until the collection of frequent patterns becomes empty.
\end{description}

\begin{algorithm}[ht]
\small
\caption{GONPM+}
\label{alg:GONPM+}
\LinesNumbered
\KwIn{\textit{SDB}, \textit{minsup}, gap constraints \textit{gap}, the decay factors $\{f_2$, $f_3$, $\dots$, $f_n\}$}.
\KwOut{Negative frequent patterns stored in \textit{F}.}

    initialize traverse \textit{SDB}, get $\Sigma$, 
    and store frequent positive patterns with length one in $F_1$; $len$ $\gets 2$;

    initialize $F_2$ $\gets$ \textbf{FindONP2(\textit{SDB}, \textit{minsup}, \textit{gap}, $F_1$, $\Sigma$)};   

    \While{\rm $F_{len}$$ \neq$ \text{null}}{
        \textit{cand} $\gets$ \textbf{PatternJoin($F_{len}$, \textit{len})};\\ 
        \tcp{\textnormal{Generate all positive and negative patterns patterns with length $\textbf{len}$ \textbf{+} $\textbf{1}$};}
        
         $\textit{minsup}_{new} \gets \textit{minsup} \times f_{len+1}$;\\
         \tcp{\textnormal{Dynamically adjust the threshold based on \textit{len};}}
                
        \For{\rm each positive candidate pattern \textbf{p} in \textit{cand} }{
        \textit{sup}(\textbf{p}, \textit{SDB}) $\gets$ \textbf{MatchDB(\textit{SDB}, \textbf{p})};\\
        
            \If{\rm \textit{minsup} $\leq$ \textit{sup}(\textbf{p}, \textit{SDB}) }{
                $F_{len + 1}$ $ \gets$ $F_{len + 1}$$ \cup$  \textbf{p};            
                Prune \textbf{p};
            }
            
            \Else{
                prune \textbf{p} and its corresponding negative sequence patterns;
                \tcp{\textnormal{Prune strategy};}
            }
        }
        $F_{len + 1}$ $ \gets$ $F_{len + 1}$ $\cup$ \textbf{FindFrequent(\textit{SDB}, \textit{minsup}, \textit{cand})};\\
        \tcp{\textnormal{Filter the frequent negative patterns from \textit{cand}};}
        
        \textit{len} $\gets$ \textit{len} + 1;\\
    }
    \textbf{return} $F$
\end{algorithm}

\textbf{Complexity analysis}. (1) The time complexity of the GONPM+ algorithm involves several parameters: $l$ represents the number of candidate patterns, $m$ denotes the maximum length of such patterns, and $n$ stands for the length of \textit{SDB}. The time complexity is mainly composed of generating candidate patterns and mining frequent patterns: the former is $O$($l^2$), while for the latter, since the \textit{MatchDB} algorithm has a time complexity of $O$($m$ $\times$ $n$) (a nettree contains $m$ levels with no more than $n$ nodes per level), mining frequent patterns is $O$($l$ $\times$ $m$ $\times$ $n$), leading to the overall time complexity of GONPM+ being $O$($l$ $\times$ $m$ $\times$ $n$ + $l^2$). (2) The space complexity of GONPM+ comprises two key components: candidate pattern generation and support computation. The space complexity of candidate patterns is $O$($m$ $\times$ $l$), since the number of candidate patterns is $l$ and the number of frequent patterns does not exceed $l$. The space complexity of the \textit{MatchDB} algorithm is $O$($m$ $\times$ $n$). Therefore, the space complexity of the GONPM+ algorithm is $O$($m$ $\times$ ($l$ + $n$)).

\textbf{Example}: Suppose that we have a sequence \textbf{S} = $\langle$$s_1$$ s_2$$ s_3$$s_4$ $s_5$$s_6$$s_7$$s_8$$s_9$$s_{10}$$s_{11}$$\rangle$ = $\langle$$\text{AACACCTCAAG}$$ \rangle$, \textit{gap} = $[M, N]$ = [0, 2], \textit{minsup} = 2, and $\Sigma$ = \{\text{A, C, G, T}\}, decay factors (0.9,0.5). \textbf{Step 1:} Iterate through the sequence to count the occurrences of each character. We obtain $F_1$ = \{$A, C$\}, since the occurrence counts of characters A and C are greater than \textit{minsup} = 2. \textbf{Step 2}: Generate positive candidate patterns with length two. Thus, we obtain $A[0,2]A$, $A[0,2]C$, $C[0,2]A$, $C[0,2]C$, since their support values are all greater than 1.8. Next, we generate negative candidate patterns based on these patterns, resulting in a total of 16 candidate patterns. However, only five of them ($A[0,2]\neg GC$, $A[0,2]\neg TC$, $C[0,2]\neg GA$, $C[0,2]\neg CC$, and $C[0,2]\neg GC$) meet the support requirement. Thus, the patterns in $F_2$ are the set of these positive and negative frequent patterns. \textbf{Step 3:} Generate positive and negative candidate patterns with length three using pattern join. At this point, \textit{minsup} is adjusted to $2 \times 0.5$ = 1. That is to say, the support of patterns in $F_3$ is 1. There are 6 positive candidate patterns in total, including $A[0,2]A[0,2]C$, $A[0,2]C[0,2]A$, $A[0,2]C[0,2]C$, $C[0,2]A[0,2]C$, $C[0,2]A[0,2]A$, and $C[0,2]C[0,2]A$. A total of 30 negative frequent patterns are obtained, including $A[0,2]C[0,2]\neg CC$, $A[0,2]\neg GC[0,2]\neg CC$, etc. $F_3$ stores these frequent and negative patterns. \textbf{Step 4}: Repeat the steps in Step 3, with \textit{minsup} set to 1 at this point, until the set of frequent patterns is empty. Then the algorithm terminates.

\subsection{Classification through discovered negative frequent patterns}

This step involves utilizing the extracted negative pattern features for virus classification. This step can be divided into two phases: the training phase and the testing phase. In the training phase, we construct appropriate ML-based classifiers. In the testing phase, the objective is to evaluate the constructed models to assess their performance in terms of the final classification results. Multi-class (MC) classification is conducted in this study, which assigns each genomic sequence a label corresponding to its unique virus type, requiring the discrimination of each virus. We utilized eight standard ML algorithms, specifically: (1) Logistic Regression (\textit{LR}), (2) Support Vector Machine (\textit{SVM}), (3) Decision Tree (\textit{DT}), (4) Random Forest (\textit{RF}), (5) k-Nearest Neighbors (\textit{kNN}), (6) Naive Bayes  (\textit{NB}), (7) Multilayer Perceptron (\textit{MLP}), and (8) Gradient boosting machine (\textit{GBM}). The performance of these classifiers is evaluated using six metrics \cite{sokolova2009systematic}: (1) Accuracy (\textit{ACC}), (2) Precision (\textit{P}), (3) Recall (\textit{R}), (4) F1 score (\textit{F1}), (5) Area Under the Curve (\textit{AUC}) and (6) Area Under the Precision-Recall Curve(\textit{AUPRC}). These six metrics are defined as follows:
\begin{equation}
    \textit{ACC} = \frac{TP + TN}{TP + TN + FP + FN}, \textit{Recall (R)} = \frac{TP}{TP + FN}, 
\end{equation}
\begin{equation}
   \textit{Precision (P)} = \frac{TP}{TP + FP}, \textit{F-measure} = 2 \times \frac{P \times R}{P + R},
\end{equation}
\begin{equation}
   \textit{AUC} = \int_{0}^{1} R(\textit{dFPR}), \textit{AUPRC} = \sum_{i=1}^{n} \frac{(R_i - R_{i-1}) \times (P_i + P_{i-1})}{2}
\end{equation}

Noted that \textit{TP} = true positive, \textit{TN} = true negative, \textit{FP} = false positive, and \textit{FN} = false negative. In the context of this study, \textit{TP} denotes the number of sequential patterns that are correctly identified as belonging to a specific virus type. \textit{TN} refers to the number of sequential patterns that are correctly classified as not belonging to a specific virus type. \textit{FP} represents the number of sequential patterns incorrectly identified as part of a specific virus type. \textit{FN} corresponds to the number of sequential patterns incorrectly classified as not belonging to the specific virus type. In Equation for \textit{AUC}, \textit{dFPR} refers to the derivative of the false positive rate \textit{FPR} = $\frac{FP}{FP + TN}$. $P_i$ and $R_i$ in Equation for \textit{AUPRC} represent the precision and recall values, respectively, at the $i-th$ decision threshold.

Algorithm \ref{alg: GeneNSPCla} provides the general pseudocode of the proposed GeneNSPCla approach. The RNA viral genome sequences are first converted into integer-based abstractions to facilitate pattern mining. NSPs are then extracted from the abstracted sequences using the GONPM algorithm. These patterns serve as input features for multiple machine learning classifiers, which are trained and evaluated using an 80:20 train–test split. The algorithm outputs the classification metrics, including \textit{ACC}, \textit{P}, \textit{R}, \textit{F1-score}, \textit{AUC}, and \textit{AUPRC}. Figure \ref{fig:GeneNSPCla} illustrates the workflow of the GeneNSPCla framework.

\begin{algorithm}[ht]
    \small
    \caption{GeneNSPCla}
    \label{alg: GeneNSPCla}
    \LinesNumbered
    \KwIn{Genome sequence corpus (\textit{GSC}) of RNA viruses in  \textit{CRF}.} 
    \KwOut{Classification results including \textit{ACC}, \textit{P}, \textit{R}, \textit{F1}, \textit{AUC}, and \textit{AUPRC}.}
    
    \tcp{\upshape\rmfamily \textbf{Step 1:} Convert sequences into the needed abstraction }
    
    abstraction $\gets$ Convert \textit{GSC} to integer-based representation;
    
    \tcp{\upshape\rmfamily \textbf{Step 2:} Find negative sequential patterns (\textit{NSPs})}

    \textit{NSPs} $\gets$  Patterns are extracted by processing the abstracted \textit{GSC} using the GONPM+ algorithm;

    \tcp{\upshape\rmfamily \textbf{Step 3:} Train classifiers}
    
    \For{\rm each classifier in Classifiers}{
    train classifier with \textit{NSPs} as features using default hyperparameters;
    }
    
    \tcp{\upshape\rmfamily \textbf{Step 4:} Evaluate classifiers}
    
    \For{\rm each classifier in Classifiers}{
    evaluate the classifier using an 80:20 training: testing ratio;
    
    store metrics: \textit{ACC}, \textit{P}, \textit{R}, \textit{F1}, \textit{AUC}, and \textit{AUPRC};
    }
    
    \textbf{return} \textit{ACC}, \textit{P}, \textit{R}, \textit{F1}, \textit{AUC}, and \textit{AUPRC}
\end{algorithm}

\section{Experimental Results}  \label{sec: result}

In this section, we focus on the multi-class virus sequence classification task. This choice is motivated by the fact that our study involves eight distinct virus types, and a binary classification formulation would inevitably introduce severe class imbalance, with the negative class comprising sequences from multiple viruses and greatly outnumbering the positive class. Such an imbalance increases the risk of overfitting and can bias the learned decision boundaries, particularly when modeling heterogeneous viral populations. In contrast, a multi-class setting treats each virus class symmetrically and aims to accurately determine the specific viral category to which each sequence belongs, thereby better reflecting the practical requirement of virus identification. Although multi-class classification is inherently more challenging and often results in lower absolute accuracy, it offers greater room for methodological improvement and provides a more rigorous evaluation of discriminative capability. Indeed, as reported in related work \cite{nawaz2024exploiting}, the best-performing classifiers under similar multi-class settings achieve accuracies of only around 50\%, underscoring the difficulty and relevance of this task.

All experiments were conducted to evaluate the performance of GeneNSPCla on RNA viral genome datasets in CRF. Eight representative RNA viruses were selected from distinct taxonomic groups to ensure diversity. The positive frequent patterns were discovered using the CM-SPAM algorithm from the SPMF data-mining library\footnote{https://www.philippe-fournier-viger.com/spmf}. In contrast, negative sequential patterns were extracted with the GONPM+ algorithm. Data preprocessing and feature unification were performed in Python, utilizing libraries such as Pandas and NumPy for data manipulation and numerical computation. The training and evaluation of eight machine learning classifiers were carried out using the scikit-learn library\footnote{https://scikit-learn.org}. To reduce classifier sensitivity to hyperparameters and ensure a fair comparison, GridSearchCV was employed to optimize key hyperparameters under the same cross-validation setting, but the resulting performance differences were marginal. All experiments were conducted on a workstation equipped with an Intel Core i7 processor, 64 GB of RAM, and the Windows 11 operating system. The parameters of the classifiers and algorithms are shown in Table \ref{tab:hyperparameters}. The source code and datasets are available at GitHub\footnote{https://github.com/zhuwenxi317/GeneNSPCla}.

\begin{table*}[ht]
\centering
\caption{Parameters of the eight ML classifiers and three algorithm settings.}
\label{tab:hyperparameters}
\begin{tabular}{p{2cm} p{12.5cm}}
    \hline
    Classifier & Parameters \\ \hline
    LR   & solver: 'lbfgs', C: 1.0, max\_iter: 1000 \\
    RF   & n\_estimators: 200, criterion: 'gini', max\_depth: None, min\_samples\_split: 2;\\
            &min\_samples\_leaf: 1, n\_jobs: -1 \\
    kNN  & n\_neighbors: 5, weights: 'distance', metric: 'minkowski' \\  
    NB  & Default parameters \\
    SVM  & kernel: 'rbf', C: 1.0, gamma: 'scale', probability: True \\
    MLP  & hidden\_layer\_sizes: (100,), activation: 'relu', solver: 'adam';\\
            &learning\_rate\_init: 0.001, max\_iter: 300\\
    DT   & criterion: 'gini', max\_depth: None, min\_samples\_split: 2, min\_samples\_leaf: 1 \\
    GBM  & n\_estimators: 100, learning\_rate: 0.1, max\_depth: 3 \\ \hline  
    Algorithm & Parameters \\ \hline
    CM-SPAM  & Min pattern length: 6, max pattern length: 9, Required items: 1, 2, 3, 4, max gap: 1 \\
    ONP-Miner & Gap constant: [0,3] \\  
    GONPM  & Gap constant: [0,3], min length: 3, ratio: 1.3 \\
    GONPM+ & Gap constant: [0,3], $(f_2, f_3, \dots, f_n)$ = (0.9,0.85,0.75,0.65)\\ \hline   
\end{tabular}
\parbox{\linewidth}{\small Regarding the parameters of the CM-SPAM algorithm, this indicates that the discovered patterns with lengths ranging from 6 to 9 are required to contain the four nucleotide bases (A, C, G, T), and the gap constraint is set to [0, 1]. Regarding the parameters of the GONPM algorithm, when $\textit{len} \geq 3$, the support is reduced to $1/1.3$ times its original value.}
\end{table*}

\subsection{Frequent patterns}

\begin{figure*}[!h]
	\centering
	\includegraphics[clip,scale=0.45]{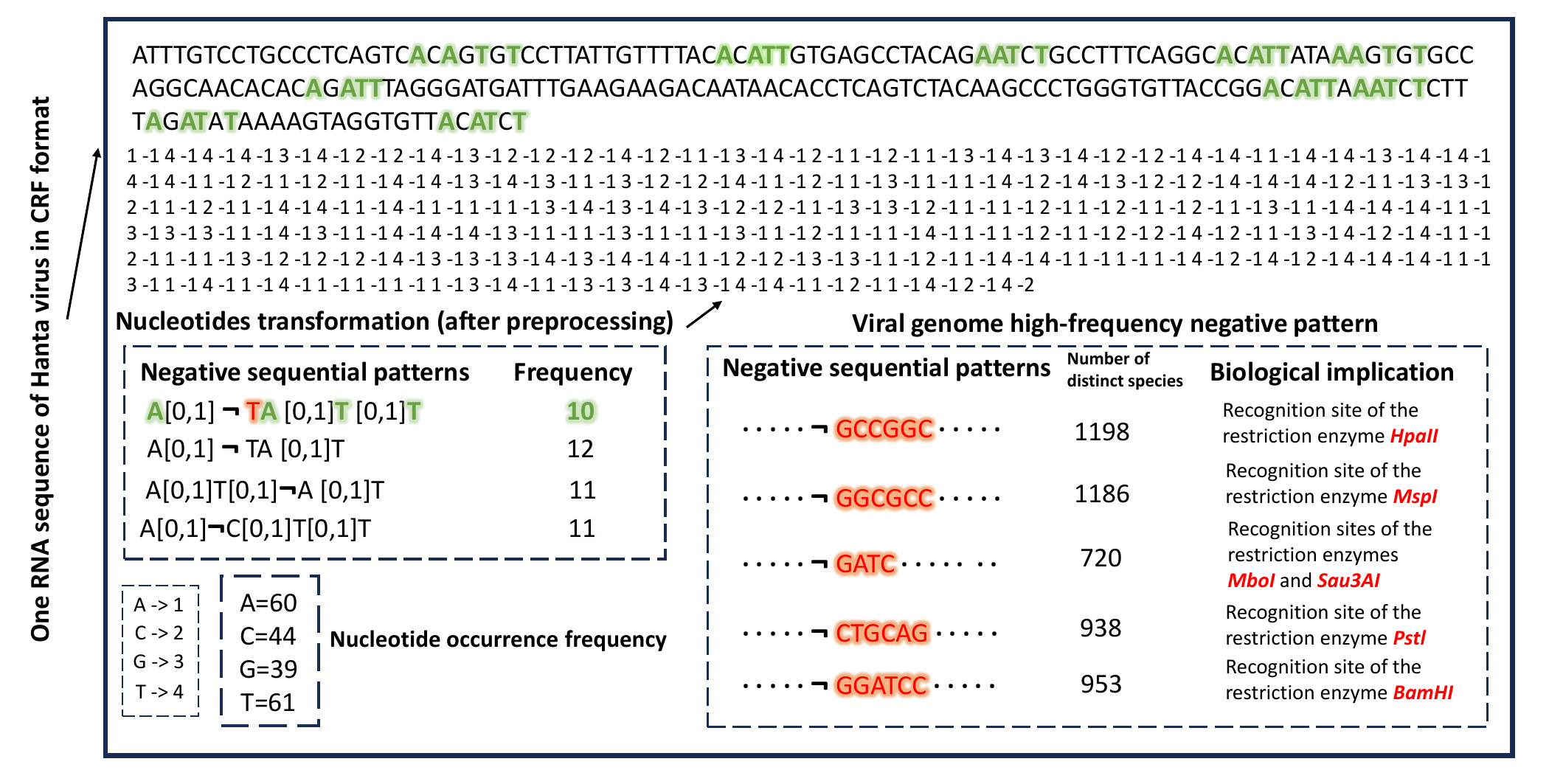}
	\caption{Example of a CRF-encoded Hanta virus sequence and the corresponding negative sequential patterns. The upper part shows the original RNA sequence and its numerical encoding. The lower-left panel displays frequent negative sequential patterns identified by GONPM+, and the lower-right panel provides illustrative biological interpretations of such negative patterns reported in the literature.}
    \label{fig:sample in sequence}
\end{figure*}

\begin{figure}[!h]
	\centering
	\includegraphics[clip,scale=0.3]{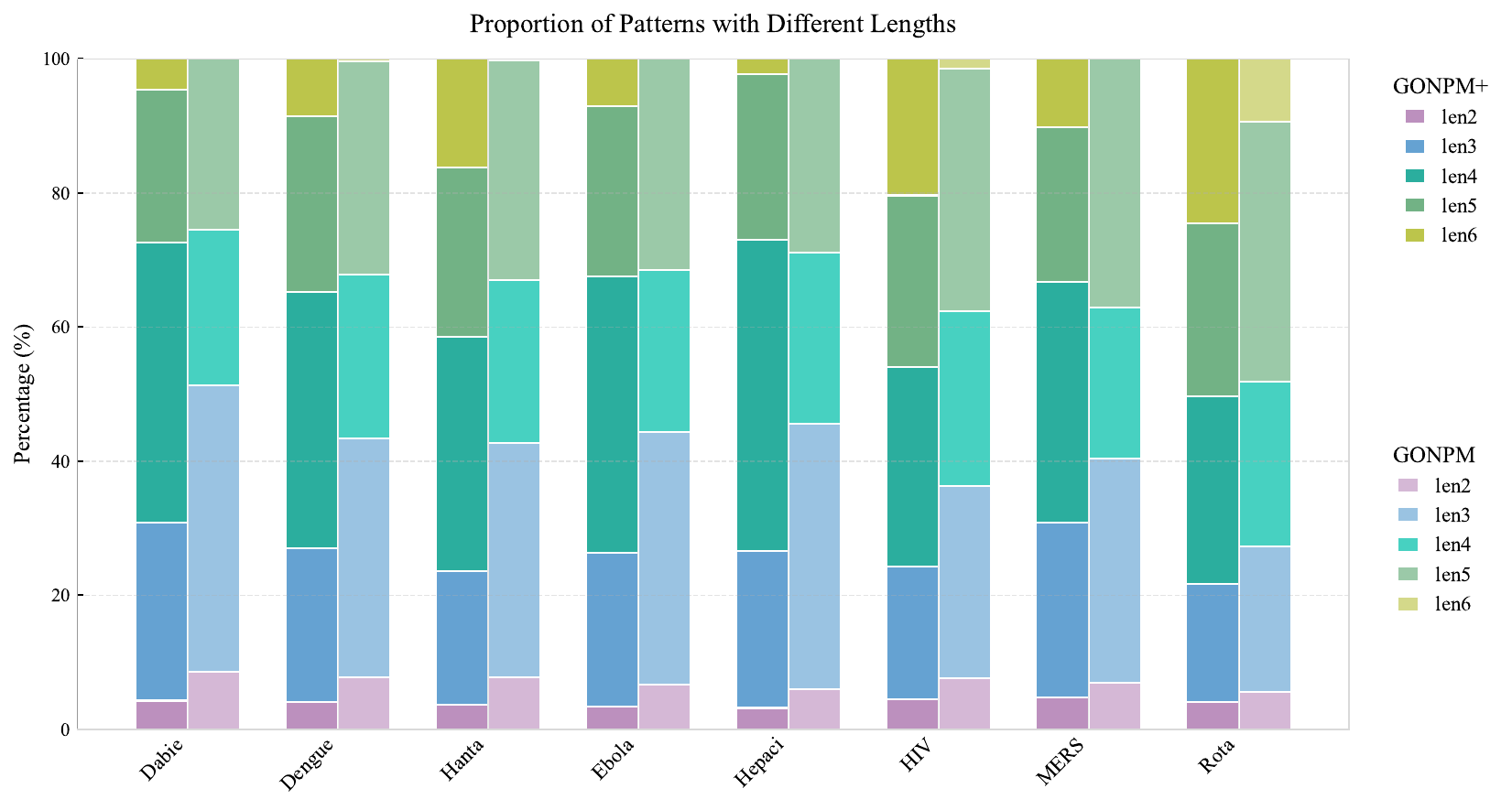}
	\caption{Proportional distribution of pattern lengths among the features used for classification, obtained using the GONPM and GONPM+ algorithms under identical threshold parameters. The darker bars on the left represent the results produced by GONPM+, while the lighter bars on the right correspond to those obtained by GONPM. For each column, the stacked segments from bottom to top denote the number of patterns with lengths 2, 3, 4, 5, and 6, respectively.}
    \label{fig:pattern propotion}
\end{figure}

\begin{figure}[!h]
	\centering
	\includegraphics[clip,scale=0.28]{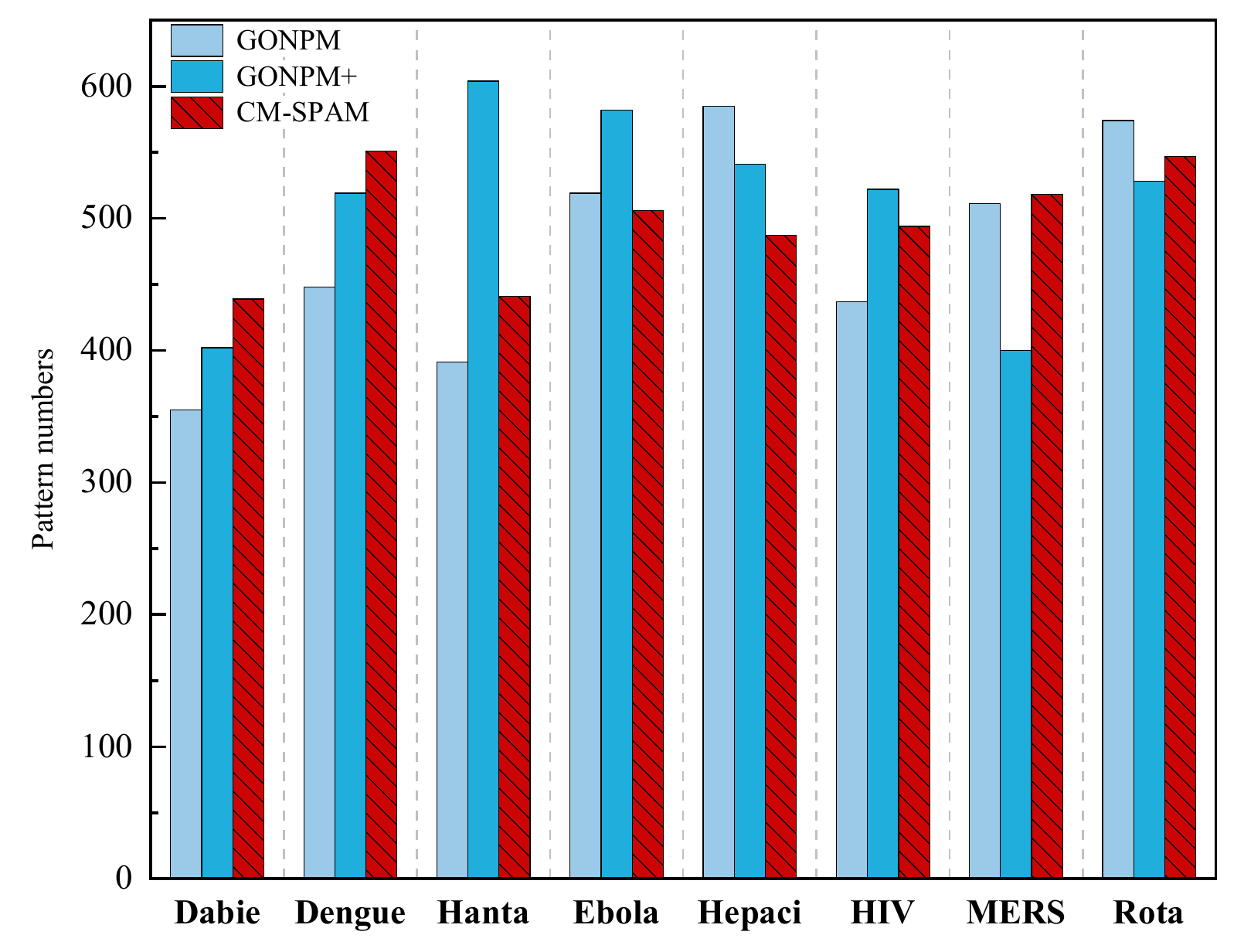}
	\caption{The number of features discovered by the GONPM, GONPM+, and CM-SPAM algorithms after processing eight virus types, which are used as input for classifier training. For each algorithm, the number of extracted patterns per virus was controlled within the range of 300–600 to maintain a comparable feature scale across classes and to avoid potential imbalance-related issues during classification.}
    \label{fig:Pattern number}
\end{figure}

\begin{figure*}[!h]
	\centering
	\includegraphics[clip,scale=0.45]{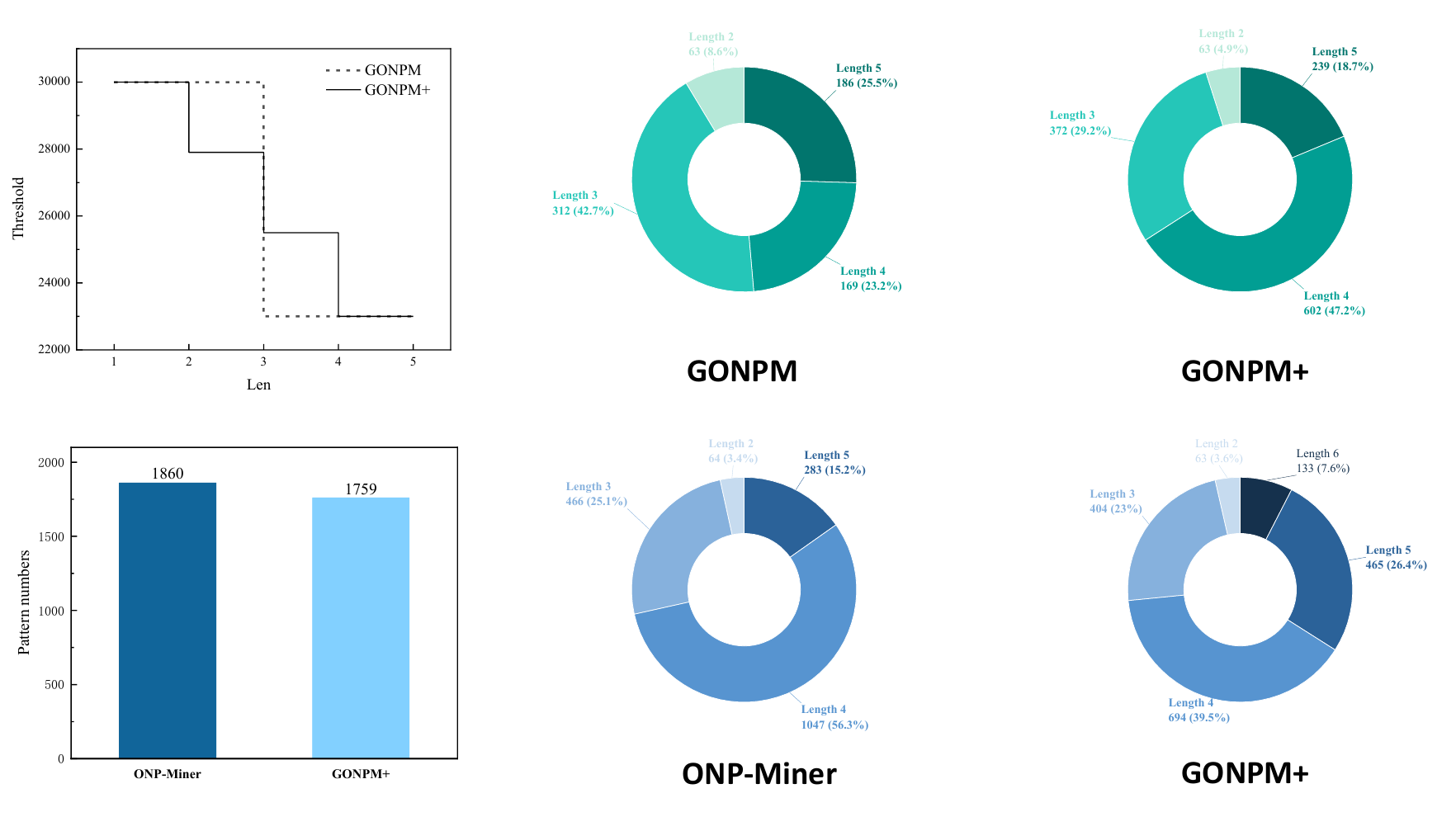}
	\caption{The upper and lower panels illustrate the effect of GONPM+ on increasing the proportion of long negative patterns under controlled experimental settings. In the upper panel, the final decay threshold is kept identical across methods, whereas in the lower panel, the total number of discovered patterns is constrained to be comparable. In both panels, the four colors from light to dark represent patterns of lengths 2, 3, 4, and 5, respectively.}
    \label{fig:proportion}
\end{figure*}

To intuitively illustrate the negative sequential patterns discovered by our method, Figure \ref{fig:sample in sequence} presents a representative example mined from a 204-nt RNA sequence of Hanta virus. The upper nucleotide sequence and the encoded numeric sequence in the middle correspond to the same viral genome, where the encoding procedure is described in detail in Section \ref{def:Database and processing}.
The green-marked bases in the sequence indicate the occurrences of a representative negative sequential pattern, $\text{A}[0,1]\neg\text{TA}[0,1]\text{T}[0,1]\text{T}$, which appears ten times in this sequence. This pattern explicitly characterizes the absence of the specific nucleotide T between the surrounding nucleotides, rather than the mere presence of frequent motifs, thereby capturing absence-based constraints in the viral genome organization.

As illustrated in the lower-right panel of Figure \ref{fig:sample in sequence}, we refer to the biological interpretation of absent or underrepresented subsequence patterns reported by Koulouras et al. \cite{koulouras2021significant}. Their study shows that negative patterns in viral genomes are not random artifacts but reflect conserved evolutionary adaptation strategies arising from long-term virus–host interactions. In particular, many recurrent negative patterns correspond to short palindromic sequences of 4–6 nucleotides that serve as recognition sites for host restriction enzymes and are therefore selectively avoided by viruses to reduce genome cleavage. A representative example is the subsequence GCCGGC, which is the recognition site of the restriction enzyme \textit{HpaII} and has been reported to be absent in 1,198 viral species, indicating a widespread and conserved avoidance mechanism across viral taxa. In addition, other absent or underrepresented sequences have been associated with immune evasion via molecular mimicry, while the systematic avoidance of long repetitive or GC-rich palindromic sequences helps prevent unfavorable secondary structure formation, thereby supporting efficient viral replication and translation.

\subsubsection{Pattern numbers and length composition}

Figure \ref{fig:pattern propotion} illustrates the proportions of patterns of different lengths among the frequent negative patterns discovered by the GONPM and GONPM+ algorithms, respectively. As can be observed from the figure, GONPM+ is capable of increasing the proportion of longer patterns and has significantly boosted the proportions of patterns of lengths 5 and 6. To improve the quality and discriminative power of the extracted features, patterns of lengths 2 and 3 were removed prior to classification for all algorithms under comparison, ensuring a fair and consistent feature selection criterion across different pattern-mining methods. Short patterns are often overly generalized and carry limited biological information, as they tend to occur frequently across different viral genomes and thus contribute minimally to distinguishing between viral species. Indeed, previous k-mer-based genome-signature studies have shown that very small k values fail to capture species-specific or taxon-specific signals, because the k-mer space is too shallow and dominated by background biases \cite{moeckel2024survey}. By excluding these short patterns with low information content, the resulting set of patterns becomes more specific and biologically meaningful.

We restricted the number of feature sequences extracted for each virus to the range from 300 to 600 to maintain a comparable feature scale across virus classes. This design choice helps prevent class imbalance at the feature level and ensures that no virus dominates the training process due to an excessive number of patterns. As a result, the resulting feature representations remain balanced and suitable for fair classifier training. The final numbers of feature sequences for different viruses are reported in Figure \ref{fig:Pattern number}.

\subsubsection{Reasons for longer patterns proportion}

In addition, through controlled experiments, we further demonstrate that GONPM+ increases the proportion of longer patterns compared to GONPM and ONP-Miner (shown in Figure \ref{fig:proportion}). In the upper panel, which compares GONPM+ with GONPM, the final threshold was controlled to be equal. Specifically, GONPM reduces the support threshold abruptly at the third level, whereas GONPM+ decreases it progressively across three levels. The green pie chart on the right shows that in GONPM, the most abundant pattern length is 3 (42.7\%), while in GONPM+, the dominant length shifts to 4 (47.2\%). In the lower panel, a comparison between GONPM+ and ONP-Miner is presented. By adjusting the parameters, the total number of extracted patterns was kept approximately consistent. The results show that ONP-Miner yields only 15.2\% of patterns with length 5, whereas GONPM+ increases this proportion to 26.4\%, and notably, patterns of length 6 also emerge, which cannot be obtained by ONP-Miner. It should be noted that the meanings of \textit{Len} in the upper-left line chart and \textit{Length} on the right differ. \textit{Len} represents different levels, corresponding to $F_{\text{len}}$. For example, when \text{len} = 3, $F_3$ stores patterns of various lengths, which may include lengths of 2, 3, or 4. In contrast, \textit{Length} denotes the actual length of an extracted pattern; for instance, the pattern $\text{C}\text{A}\neg\text{GT}$ has a length of 4.

The increase in the proportion of longer patterns extracted by GONPM+ is primarily due to the change in the number of candidate patterns. The negative patterns discovered by the algorithm are selected from the candidate patterns generated in the previous level. For instance, the negative patterns stored in $F_4$ are derived from the patterns stored in $F_3$, which are further expanded using the PatternJoin algorithm and then filtered based on their support values relative to the threshold. By introducing decay factors $\{f_1, f_2, \dots, f_n\}$, the threshold is progressively reduced in a stepwise manner, avoiding the explosive increase in candidate patterns seen in ONP-Miner. Unlike GONPM, where the support threshold remains unchanged in earlier levels for shorter patterns, GONPM+ causes a gradual and moderate increase in the number of patterns. This stepwise reduction in threshold allows the algorithm to incrementally mine patterns at deeper levels without generating excessive numbers of candidates, thereby facilitating the discovery of longer patterns while controlling the overall growth of candidate patterns.

\subsection{Classification results and Comparison}

\begin{table*}[htbp]
  \renewcommand{\arraystretch}{1.5}
  \centering
  \small
  \caption{Classifiers' performance across multiple evaluation metrics (in \%).}
  \label{tab:classifier_result}
  \begin{tabular}{lcccccc}
    \toprule
    Classifier       & Accuracy & Precision & Recall & F1-Score & AUC & AUPRC\\
\midrule
    LR   & $\frac{59.15\,(54.58)}{50.62\,(32.73)}$ & $\frac{59.84\,(52.64)}{49.99\,(32.82)}$ & $\frac{59.15\,(54.58)}{50.62\,(32.73)}$ & $\frac{59.16\,(53.02)}{49.65\,(29.44)}$ & $\frac{0.92\,(0.90)}{0.89\,(0.73)}$ & $\frac{0.64\,(0.59)}{0.55\,(0.30)}$ \\
    
    SVM  & $\frac{58.66\,(54.06)}{48.77\,(32.61)}$ & $\frac{60.30\,(53.36)}{48.13\,(36.05)}$ & $\frac{58.66\,(54.06)}{48.77\,(32.61)}$ & $\frac{58.46\,(53.34)}{48.68\,(31.25)}$ & $\frac{0.91\,(0.89)}{0.88\,(0.72)}$ & $\frac{0.63\,(0.57)}{0.52\,(0.28)}$ \\
    
    DT   & $\frac{54.63\,(48.82)}{42.90\,(31.96)}$ & $\frac{55.24\,(47.92)}{41.08\,(34.06)}$ & $\frac{54.63\,(48.82)}{42.90\,(31.96)}$ & $\frac{54.40\,(47.95)}{41.41\,(31.38)}$ & $\frac{0.91\,(0.87)}{0.85\,(0.71)}$ & $\frac{0.63\,(0.53)}{0.46\,(0.28)}$ \\
    
    RF   & $\frac{54.63\,(49.08)}{41.98\,(31.84)}$ & $\frac{55.94\,(47.43)}{37.57\,(34.17)}$ & $\frac{54.63\,(49.08)}{41.98\,(31.84)}$ & $\frac{54.52\,(47.67)}{39.39\,(31.32)}$ & $\frac{0.91\,(0.88)}{0.86\,(0.71)}$ & $\frac{0.64\,(0.54)}{0.48\,(0.28)}$ \\
    
    kNN  & $\frac{55.73\,(48.56)}{43.21\,(25.80)}$ & $\frac{55.84\,(49.68)}{43.05\,(25.79)}$ & $\frac{55.73\,(48.56)}{43.21\,(25.80)}$ & $\frac{55.59\,(48.86)}{42.59\,(24.82)}$ & $\frac{0.85\,(0.83)}{0.80\,(0.64)}$ & $\frac{0.53\,(0.48)}{0.41\,(0.23)}$ \\
    
    NB   & $\frac{54.76\,(51.70)}{45.06\,(29.27)}$ & $\frac{54.68\,(51.79)}{45.83\,(31.14)}$ & $\frac{54.76\,(51.70)}{45.06\,(29.27)}$ & $\frac{54.45\,(51.00)}{44.38\,(26.98)}$ & $\frac{0.90\,(0.89)}{0.86\,(0.69)}$ & $\frac{0.59\,(0.57)}{0.50\,(0.26)}$ \\
    
    MLP  & $\frac{56.95\,(51.44)}{48.46\,(33.12)}$ & $\frac{57.60\,(49.49)}{47.70\,(35.25)}$ & $\frac{56.95\,(51.44)}{48.46\,(33.12)}$ & $\frac{56.49\,(49.09)}{47.58\,(30.51)}$ & $\frac{0.92\,(0.90)}{0.88\,(0.73)}$ & $\frac{0.64\,(0.58)}{0.53\,(0.30)}$ \\
    
    GBM  & $\frac{55.85\,(51.70)}{46.91\,(32.86)}$ & $\frac{56.92\,(50.06)}{42.38\,(33.36)}$ & $\frac{55.85\,(51.70)}{46.91\,(32.86)}$ & $\frac{55.88\,(49.96)}{44.11\,(31.07)}$ & $\frac{0.92\,(0.89)}{0.88\,(0.71)}$ & $\frac{0.64\,(0.57)}{0.51\,(0.29)}$ \\
    
    \midrule
    Average & $\frac{56.02\,(51.24)}{45.99\,(31.27)}$ & $\frac{56.77\,(50.30)}{44.47\,(32.83)}$ & $\frac{56.02\,(51.24)}{45.99\,(31.27)}$ & $\frac{55.78\,(49.99)}{44.72\,(29.60)}$ & $\frac{0.90\,(0.88)}{0.86\,(0.71)}$ & $\frac{0.62\,(0.55)}{0.50\,(0.28)}$ \\
    \midrule
    AAI & $\frac{9.32}{21.8\,(79.1)}$ & $\frac{12.86}{26.0\,(70.6)}$ & $\frac{9.32}{21.8\,(79.1)}$ & $\frac{11.58}{25.3\,(89.3)}$ &$\frac{2.3}{4.7\,(26.8)}$ & $\frac{12.7}{24.0\,(121.4)}$ \\
    \bottomrule
    
  \end{tabular}
  \parbox{\linewidth}{\small
    Classifiers: LR = logistic regression, SVM = support vector machine, DT = decision tree, RF = random forest, kNN = k-nearest neighbors, NB = naive Bayes, MLP = multilayer perceptron, GBM = gradient boosting machine. The data in the table, including all evaluation metrics such as Accuracy, Precision, Recall, F1-Score, AUC, and AUPRC, follow the format $\frac{\textit{GONPM+}\,(\textit{GONPM})}{\textit{ONP-Miner}\,(\textit{CM-SPAM})}$. AAI stands for Average Accuracy Improvement, where the numerator represents the relative improvement of GONPM over ONP-Miner, and the denominator represents the relative improvement of GONPM over CM-SPAM. AAI follow the format $\frac{\textit{GONPM+}}{\textit{ONP-Miner}\,(\textit{CM-SPAM})}$.
    }
\end{table*}

\subsubsection{Classification result analysis}

Table \ref{tab:classifier_result} presents the multi-class classification performance of eight machine learning classifiers, trained on features generated by three different mining algorithms: CM-SPAM, ONP-Miner, and the proposed GONPM+. The proposed GONPM algorithm achieved an average accuracy improvement (AAI) across the evaluated classifiers compared with the baseline methods. AAI is defined as:
\begin{equation}
   \text{AAI} \;=\; \frac{1}{n}\sum_{i=1}^{n}\frac{A_{\text{GONPM+},i}-A_{\text{baseline},i}}{A_{\text{baseline},i}}\times 100\% 
\end{equation}
where \(A_{\text{GONPM+},i}\) denotes the accuracy of GONPM+ with the \(i\)-th classifier, \(A_{\text{baseline},i}\) denotes the accuracy obtained by a baseline pattern mining algorithm (e.g., ONP-Miner or CM-SPAM) with the same classifier, and \(n\) is the number of classifiers used in the comparison (here \(n=8\)). This metric reports the mean relative gain of GONPM+ over the baseline across all classifiers. Overall, the results clearly demonstrate that GONPM+ consistently achieves the highest classification performance across all classifiers. Among the eight evaluated classifiers, the SVM achieved the best performance on GONPM+, with an accuracy of 58.66\%. In comparison, the highest accuracies achieved by GONPM, ONP-Miner, and CM-SPAM were 54.58\%, 50.62\%, and 32.73\%, respectively. The average classification accuracy across all classifiers was 56.02\% for GONPM+, compared with 51.24\%, 45.99\%, and 31.27\% for the other algorithms. Moreover, GONPM+ outperformed all baselines not only in accuracy but also across other key evaluation metrics, confirming its superior robustness and generalization capability.

\begin{figure}[h]
	\centering
	\includegraphics[clip,scale=0.3]{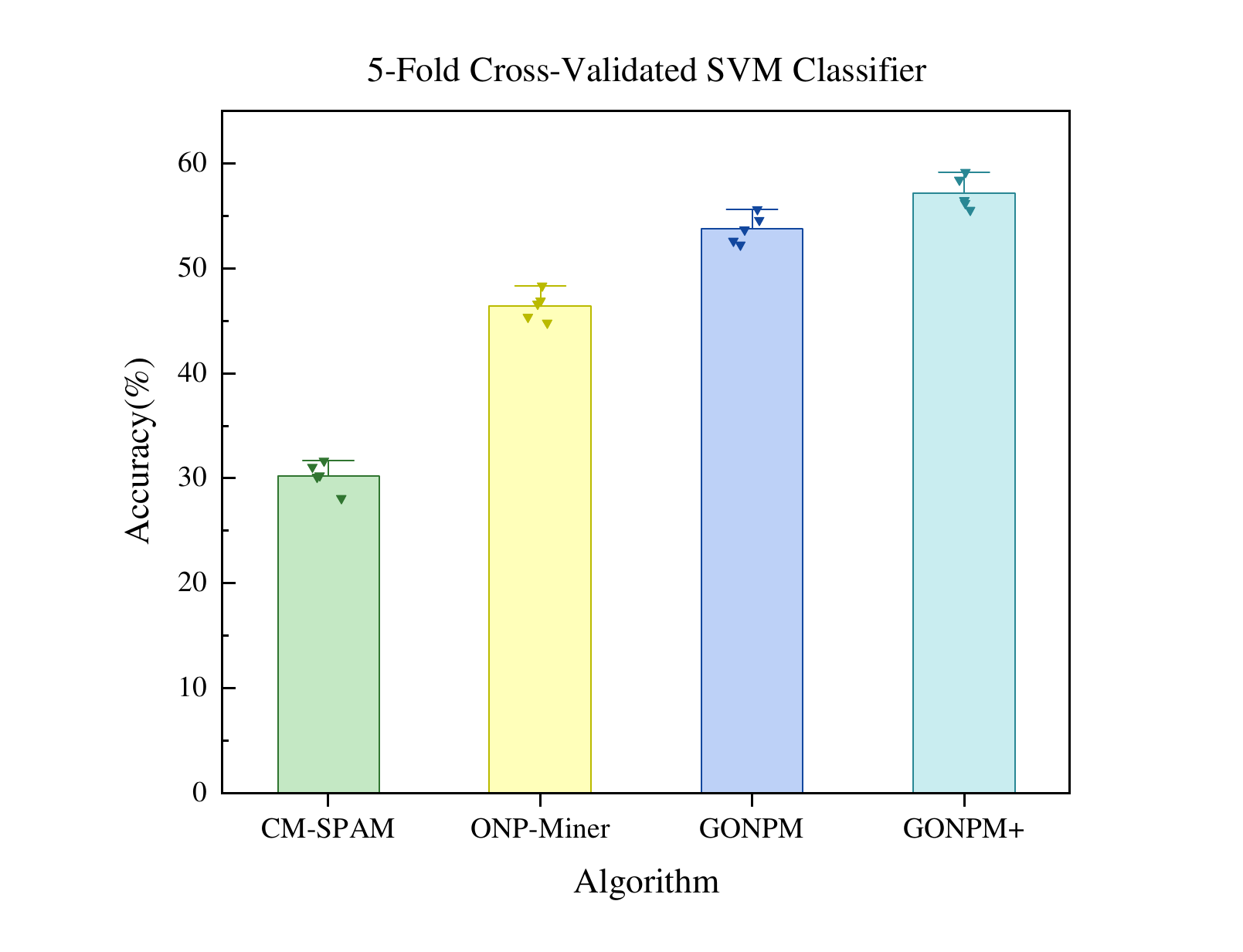}
	\caption{Accuracy comparison of SVM classifiers trained on features generated by four sequence pattern mining algorithms (CM-SPAM, ONP-Miner, GONPM, and GONPM+) under 5-fold cross-validation. The height of each bar represents the average classification accuracy (\%), error bars indicate the range of accuracy fluctuations across 5 validation folds, and triangle points denote the accuracy of each individual fold.}
    \label{fig:5-Flod}
\end{figure}

\begin{figure*}[!h]
	\centering
	\includegraphics[clip,scale=0.55]{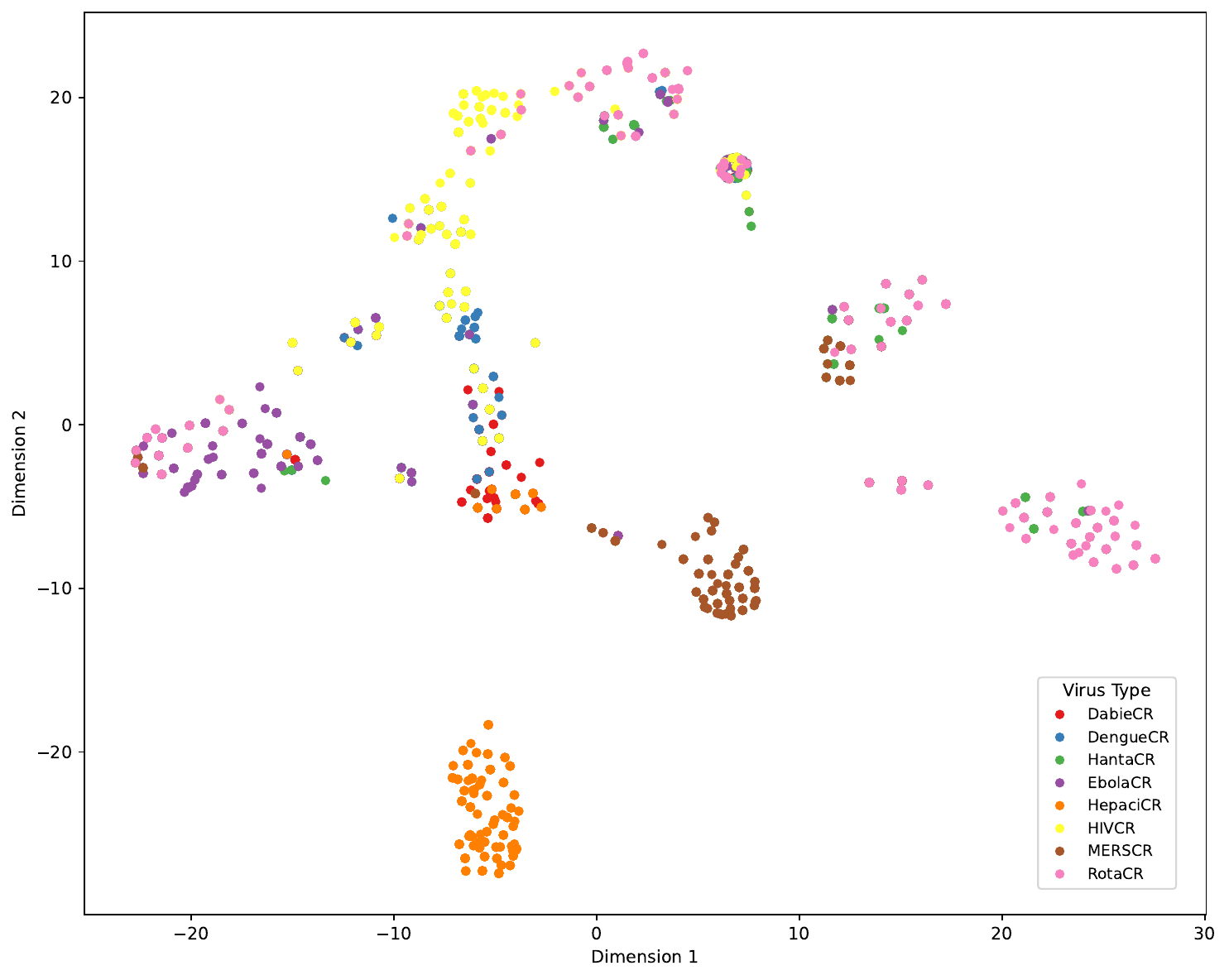}
	\caption{Visualization of the multi-class classification results of eight RNA viruses using t-SNE based on features extracted by the GONPM+ algorithm. Each color represents a distinct viral species, and the spatial separation of clusters indicates the discriminative effectiveness of the discovered negative sequential patterns. The "CR" suffix related to viruses denotes the viral Coding Region Form.}
    \label{fig:tSNE}
\end{figure*}

\begin{figure*}[!h]
	\centering
	\includegraphics[clip,scale=0.6]{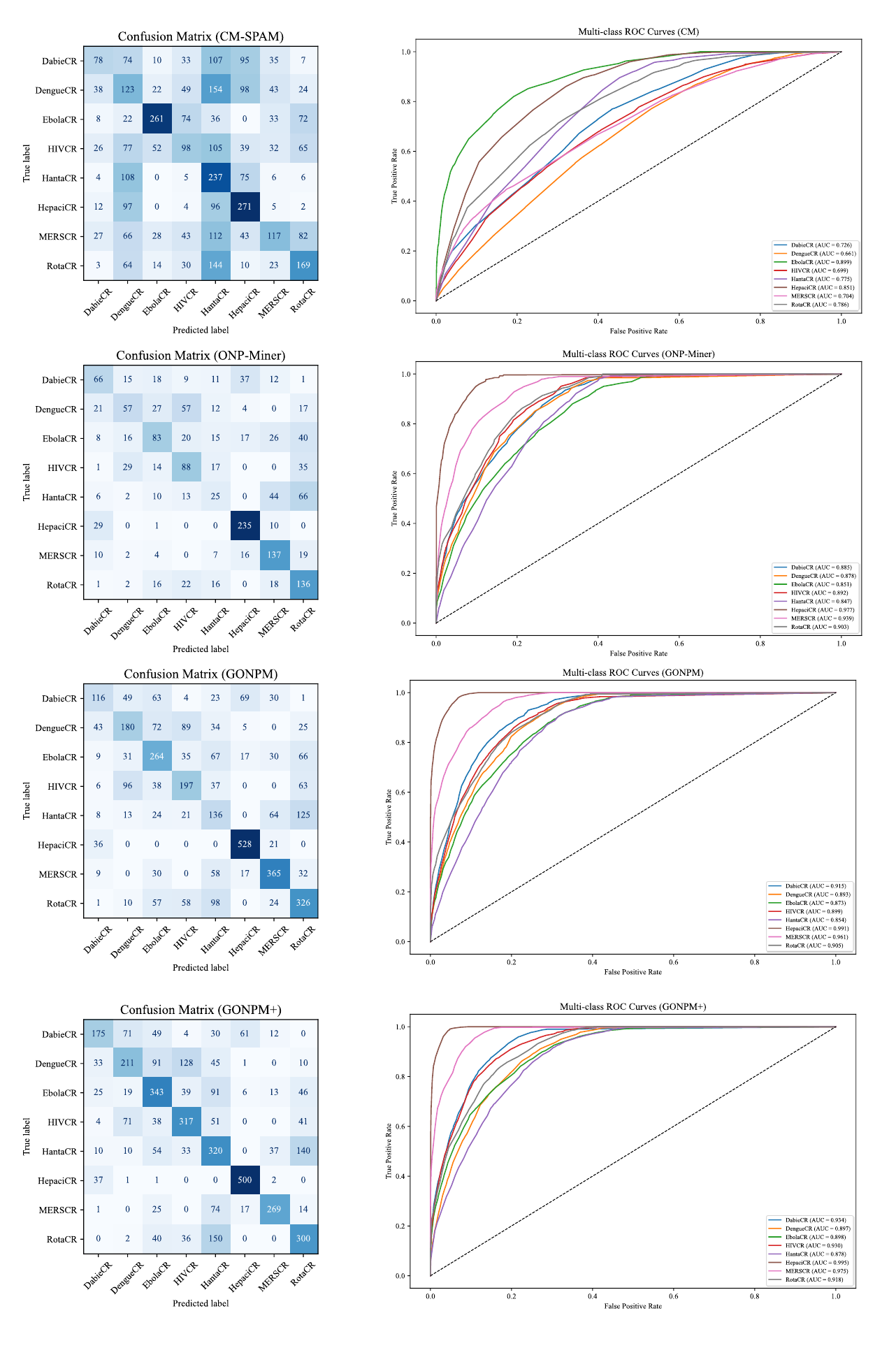}
	\caption{Confusion matrices and ROC curves obtained using the CM-SPAM, ONP-Miner, GONPM, and GONPM+ algorithms, respectively, based on the RF classifier. The results are arranged from top to bottom in the same order. The confusion matrices reflect the ability of the RF classifier to distinguish among eight different virus types, whereas the ROC curves highlight the overall discriminative performance achieved by different pattern-mining strategies.}
    \label{fig:Classification results}
\end{figure*}

Figure \ref{fig:5-Flod} presents the accuracy results under 5-fold cross-validation, demonstrating the robustness of the proposed method. Figure \ref{fig:tSNE} shows the t-SNE visualization of classification results for eight RNA viruses using GONPM+-extracted features, revealing distinct inter-species separability. Figure \ref{fig:Classification results} illustrates the confusion matrices and ROC curves, obtained from the Random Forest (RF) classifier for the eight viral classes, using three feature-extraction algorithms. As shown, the GONPM+-based model achieves the highest classification accuracy, with fewer misclassifications and higher recall across nearly all virus categories. Interestingly, we found that both positive and negative pattern mining algorithms achieved favorable results in identifying Hepaci viruses, indicating that this virus is highly distinguishable. Additionally, in the identification of Hanta and Dengue viruses, the results obtained by both algorithms were unsatisfactory, suggesting that these viruses lack sufficient distinguishability. For the improved GONPM algorithm, the probability that the predicted label equals the true label is consistently the highest in the prediction of each virus. These visual results confirm that GONPM+ achieves more robust multi-class classification performance.

\subsubsection{Cause Analysis}

Compared with GONPM, the proposed GONPM+ algorithm achieves superior classification performance primarily due to its finer-grained control over the minimum support threshold. By introducing decay factors \{$f_2$, $f_3$, $\dots$, $f_n$\}, GONPM+ gradually reduces the support value at each mining level instead of applying a single abrupt change as in GONPM. This progressive adjustment allows the algorithm to retain more potentially informative longer patterns while still filtering out redundant or noisy candidates. These longer patterns constitute the major portion of the extracted feature sequences and contain richer information, thereby contributing to more effective and accurate viral classification results. On average, GONPM+ improves the overall classification accuracy by approximately 4.78\% compared to GONPM.

Compared with the positive pattern mining algorithm CM-SPAM, the accuracy has improved by nearly 25\%. The potential reasons for this improvement are as follows: Positive pattern mining focuses solely on the presence of recurring motifs, which often appear across multiple viral species, and may therefore lack sufficient specificity. Negative pattern mining retains the information of positive frequent patterns, as negative patterns are derived from positive frequent patterns through pruning strategies. Furthermore, each sequence of positive patterns has dimension 4, as it contains only $A$, $C$, $G$, and $T$. By contrast, negative patterns have eight unique tokens, with the addition of $\neg A$, $\neg C$, $\neg G$, and $\neg T$ to the positive patterns, resulting in each sequence having a dimension of 8. This enhances the uniqueness of the patterns, leading to more robust and accurate multi-class viral classification results.

\subsection{Comparison with SOTA approaches}

In this study, we compare the proposed method with GenoAnaCla \cite{nawaz2024exploiting}, as both approaches adopt data mining–based strategies for viral genome classification. The two methods are benchmarked on the same dataset with an identical preprocessing and encoding scheme, ensuring a fair comparison in which the primary difference lies in the pattern-mining algorithm itself. Under these controlled settings, the proposed method achieves improved classification performance, as shown in Table~\ref{tab:comparison}. In contrast, direct comparison with deep learning–based models is not pursued, since their performance is highly dependent on dataset scale and preprocessing choices, making such benchmarks less interpretable under the current experimental conditions.

\begin{table}[h]
    \centering
    \small
    \caption{Comparison of GeneNSPCla with SOTA approaches.}
    \label{tab:comparison}
    \begin{tabular}{p{1.2cm} p{3cm} p{3cm}} \hline
    \textbf{Approach} & \textbf{GenoAnaCla} \cite{nawaz2024exploiting} & \textbf{GeneNSPCla} \\ \hline
Datasets & 8 virus   &8 virus \\

Methods & Positive pattern mining + 8 types of ML-based models & Negative pattern mining + 8 types of ML-based models \\

Algorithm & CM-SPAM & GONPM+, GONPM \\

$O_t$ & $O$($n$ $\times$ $m^2$) &$O$($l$ $\times$ $m$ $\times$ $n$ + $l^2$)\\

$O_{space}$ & $O$($n$ $\times$ $m$ + $l$) & $O$($m$ $\times$ ($l$ + $n$))\\

Result     & ACC: 31.27& ACC: \textbf{56.02}\\
           & AUC: 0.71 & AUC: \textbf{0.91}\\      \hline
\end{tabular}
\end{table}

From a computational efficiency perspective, considering $n$ as the sequence database length, $m$ as the maximum candidate pattern length, and $l$ as the number of candidate patterns, the CM-SPAM algorithm used in GenoAnaCla has a time complexity of $O(n \times m^2)$, which grows quadratically with $m$ and can lead to rapidly increasing runtime for long-pattern datasets such as viral genomes. In contrast, the GONPM+ algorithm employed in GeneNSPCla exhibits a time complexity of $O(l \times m \times n + l^2)$, where the dependence on $m$ is linear, and the number of candidate patterns $l$ is effectively controlled through pruning, resulting in more gradual computational growth in practice. Regarding space complexity, CM-SPAM requires $O(n \times m + l)$ memory, while GONPM+ requires $O(m \times (l + n))$; since both share the dominant term $m \times n$, their memory consumption remains comparable when $l$ is not excessive. Overall, GeneNSPCla avoids the quadratic dependence on pattern length inherent in CM-SPAM without introducing substantial additional memory overhead, making it more suitable for viral sequence analysis.

\section{Conclusion} \label{sec: conclusion}
\subsection{Result analysis and conclusion}

In this study, we propose a novel framework termed GeneNSPCla, an enhanced negative sequential pattern mining framework for RNA virus classification, together with an improved mining algorithm, GONPM+, which dynamically adjusts the minimum support threshold during pattern generation to extract longer and more informative negative patterns compared with the conventional ONP-Miner algorithm. Experimental results demonstrate that features derived from GONPM+ consistently improve classification performance across multiple machine learning classifiers, achieving average accuracy gains of 10.03\% and 23.16\% over ONP-Miner and the positive pattern mining algorithm CM-SPAM, respectively. While these results validate the methodological effectiveness of the proposed approach, we acknowledge that certain limitations remain and that further improvements are possible, which are discussed in the following sections. Overall, the findings suggest that negative sequential patterns provide a complementary and more discriminative feature representation than positive patterns, leading to more effective classification of RNA viruses under the evaluated experimental settings. In particular, the proposed GONPM+ algorithm is better suited for mining negative patterns in multi-class RNA virus classification tasks with a limited number of virus categories, as it enables the extraction of longer negative sequential patterns through adaptive support adjustment. These longer patterns enrich the feature space and contribute to improved classification performance compared with conventional positive pattern mining and earlier negative pattern approaches.

\subsection{Limitations and future directions}

Despite the encouraging results, several limitations remain and motivate future research directions. (1) Although GeneNSPCla consistently outperforms the selected pattern-mining baselines, the absolute performance gains are moderate; future work will explore the integration of GONPM-based features with advanced learning models to further enhance classification performance. (2) GONPM+ faces scalability challenges on larger datasets, and its sensitivity to key parameters (e.g., decay factor and support threshold) has not been fully explored; future studies will focus on improving computational efficiency and conducting systematic parameter sensitivity analyses to refine the dynamic threshold adjustment strategy. (3) The current evaluation relies on a limited-scale dataset with 200 sequences per virus class and eight RNA virus families; future work will extend validation to larger, more diverse datasets, including closely related viral strains and real-world metagenomic mixtures, to better assess robustness and generalizability. (4) While negative sequential patterns show strong discriminative capability, their biological interpretation has not yet been examined; future research will aim to associate representative negative patterns with known viral functions or evolutionary mechanisms to enhance biological relevance and interpretability.

\section*{Acknowledgment}

This research was supported in part by the National Natural Science Foundation of China (No. 62272196), the Guangzhou Basic and Applied Basic Research Foundation (No. 2024A04J9971).

\section*{CRediT Authorship Contribution Statement}
\textbf{Wenxi Zhu}: Methodology, writing original draft.
\textbf{Wensheng Gan}: Review and editing, supervision.
\textbf{Zhenlian Qi}: Review and editing.

% \bibliographystyle{cas-model2-names}
\bibliographystyle{apalike}
\bibliography{main.bib}

\end{document}